\documentclass[prd,amsmath,amssymb,superscriptaddress,floatfix,nofootinbib,10pt]{revtex4}
\usepackage{times}
\usepackage{amssymb,amsbsy,amsmath,amsfonts}
\usepackage{graphicx}
\usepackage{float}
\usepackage{color}
\usepackage{orcidlink}
\usepackage{morefloats}
\usepackage{rotating}
\usepackage{srcltx}
\usepackage{slashed}
\usepackage{multirow}
\usepackage{verbatim}
\usepackage{hyperref}
\usepackage{tabularx}
\usepackage{bm}
\allowdisplaybreaks[4]

\usepackage{bbding}
\usepackage{threeparttable}

\usepackage{mathrsfs}

\DeclareUnicodeCharacter{2212}{\ensuremath{-}}

\newcommand{\PreserveBackslash}[1]{\let\temp=\\#1\let\\=\temp}
\newcolumntype{C}[1]{>{\PreserveBackslash\centering}p{#1}}
\newcolumntype{R}[1]{>{\PreserveBackslash\raggedleft}p{#1}}
\newcolumntype{L}[1]{>{\PreserveBackslash\raggedright}p{#1}}

\begin{document}

\title{Revisiting the  $\Lambda_c^+\rightarrow\bar{K}^0\eta p$ reaction: the role of $N^*(1535),~ N^*(1650)$ and $\Sigma(1620)$}

\author{Jing Song\,\orcidlink{0000-0003-3789-7504}}
\email[]{Song-Jing@buaa.edu.cn}
\affiliation{School of Physics, Beihang University, Beijing, 102206, China}
\affiliation{Departamento de Física Teórica and IFIC, Centro Mixto Universidad de Valencia-CSIC Institutos de Investigación de Paterna, 46071 Valencia, Spain}

\author{Melahat Bayar\, \orcidlink{0000-0002-5914-0126 }}
\email[]{melahat.bayar@kocaeli.edu.tr}
\affiliation{Department of Physics, Kocaeli Univeristy, 41380, Izmit, Turkey}
\affiliation{Departamento de Física Teórica and IFIC, Centro Mixto Universidad de Valencia-CSIC Institutos de Investigación de Paterna, 46071 Valencia, Spain}

\author{Yi-Yao Li\,\orcidlink{0009-0001-6943-4646}}
\email[]{liyiyao@m.scnu.edu.cn}
\affiliation{
State Key Laboratory of Nuclear Physics and 
Technology, Institute of Quantum Matter, South China Normal 
University, Guangzhou 510006, China}
\affiliation{Key Laboratory of Atomic and Subatomic Structure and Quantum Control (MOE), Guangdong-Hong Kong Joint Laboratory of Quantum Matter, Guangzhou 510006, China }
\affiliation{ Guangdong Basic Research Center of Excellence for Structure and Fundamental Interactions of Matter, Guangdong Provincial Key Laboratory of Nuclear Science, Guangzhou 510006, China  }
\affiliation{Departamento de Física Teórica and IFIC, Centro Mixto Universidad de Valencia-CSIC Institutos de Investigación de Paterna, 46071 Valencia, Spain}

\author{ Eulogio Oset\,\orcidlink{ https://orcid.org/0000-0002-4462-7919}}
\email[]{oset@ific.uv.es}
\affiliation{Departamento de Física Teórica and IFIC, Centro Mixto Universidad de Valencia-CSIC Institutos de Investigación de Paterna, 46071 Valencia, Spain}
\affiliation{Department of Physics, Guangxi Normal University, Guilin 541004, China}


\begin{abstract}
We perform a theoretical study of the weak decay $\Lambda_c^+ \rightarrow \bar{K}^0 \eta p$ using a coupled-channel chiral unitary approach that incorporates both pseudoscalar-baryon and vector-baryon interactions. Our framework includes contributions from both internal and external weak emission mechanisms, as well as strong final state interactions. We assume that the $N^*(1535)$ and $N^*(1650)$ resonances are dynamically generated through meson-baryon scattering and they appear as distinct structures in the $\eta p$ invariant mass distributions. {A clear peak also appears in the $\bar{K}^0 p$ invariant mass distribution around 1620~MeV, associated with the dynamically generated $\Sigma(1620)$ resonance.} Notably, this work provides the first theoretical description of the simultaneous observation of these two related $N^*$ resonances in the same meson-baryon final state.  Our results highlight the crucial role of final state interaction and the interplay between different weak decay topologies in shaping the resonance patterns. These findings offer new insights into the nature of nucleon excitations and support the interpretation of $N^*(1535)$ and $N^*(1650)$ as dynamically generated states. {Moreover, the identification of the $\Sigma(1620)$ further supports the picture of hadronic molecular structures emerging from meson-baryon interactions in the non-perturbative QCD regime.}
\end{abstract}

\maketitle

\section{Introduction} \label{sec:Intr}

Understanding the dynamics of charmed baryon decays provides valuable insight into the strong interaction in the non-perturbative  QCD regime. Among the charmed baryons, the $\Lambda_c^+$, plays a key role as a laboratory to study hadronization mechanisms and final state interactions (FSI) involving meson-baryon systems. In particular, the hadronic weak decay $\Lambda_c^+ \rightarrow \bar{K}^0 \eta p$ has attracted increasing attention. It features a relatively large branching ratio~\cite{CLEO:1995cbq,BESIII:2018ciw,ParticleDataGroup:2024cfk} and provides access to some $N^*$ states of the rich spectrum of nucleon excitations via the $\eta p$ invariant mass distributions~\cite{Belle:2022pwd}.

From a theoretical perspective, many studies have focused on quark-level mechanisms such as external and internal $W-$emission, $W-$exchange, and $W-$annihilation~\cite{Korner:1992wi, Cheng:2018hwl, Gutsche:2018utw, Gutsche:2019iac}. However, quark-level diagrams alone do not capture the strong meson-baryon interactions that can take place after hadronization. These final state interactions can significantly affect observables, including producing peaks associated with dynamically generated resonances. Therefore, effective field theories and coupled-channel methods are essential to bridge the gap between quark-level processes and hadronic observables.

A particularly interesting aspect of the $\Lambda_c^+ \rightarrow \bar{K}^0 \eta p$ decay is the formation of the $\eta p$ system in the final state, which allows the investigation of $N^*$ resonances, such as the $N^*(1535)$ and $N^*(1650)$, both with spin-parity $J^P=1/2^-$. Traditionally, these resonances were studied in pion-induced reactions~\cite{Cutkosky:1979fy, Arndt:2006bf, Workman:2012hx}, photoproduction~\cite{Anisovich:2011fc, Anisovich:2012ct, Anisovich:2013tij}, and more recently, in decays of heavy-flavored baryons~\cite{Xie:2017erh,Pavao:2018wdf,Li:2024rqb,Li:2025gvo,Li:2025msk}.
However, their internal structure is still under debate. Some works suggest they are conventional three-quark states~\cite{Liu:2015ktc,Xiao:2016dlf,Tan:2025kjk}, while others propose they are dynamically generated from meson-baryon interactions~\cite{Kaiser:1995cy, Inoue:2001ip,Hyodo:2011ur,Doring:2013glu,Mart:2013ida,Khemchandani:2013nma,Sekihara:2015gvw,Guo:2017jvc,Li:2023pjx,Liu:2025eqw}.

The idea of dynamically generated resonances is supported by several unitarized chiral models and approaches based on local hidden gauge symmetries~\cite{Oset:1997it, Lutz:2001yb, Sarkar:2004jh, Sarkar:2005ap, Hyodo:2002pk, Khemchandani:2008rk,Kaiser:1995eg,Oller:2000fj}. For instance, the $N^*(1535)$ has been interpreted as arising from the coupled-channel interactions of $\pi N$, $\eta N$, $K \Lambda$, and $K \Sigma$\cite{Kaiser:1995cy,Inoue:2001ip,Hyodo:2011ur,Li:2023pjx}. Yet, some approaches such as~\cite{Inoue:2001ip} only obtain the $N^*(1535)$, while others~\cite{Nieves:2001wt, Bruns:2010sv} also reproduce the $N^*(1650)$, depending on how off-shell effects or higher-order terms are implemented. On the other hand, the inclusion of both pseudoscalar-baryon ($PB$) and vector-baryon ($VB$) interactions allowed to generate both states simultaneously in~\cite{Garzon:2014ida, Khemchandani:2013nma}.

Experimentally, the $N^*(1535)$ was observed in the $\eta p$ invariant mass distribution of $\Lambda_c^+ \rightarrow \bar{K}^0 \eta p$ in the Belle measurement~\cite{Belle:2022pwd}, confirming earlier theoretical predictions~\cite{Xie:2017erh}. In Ref.~\cite{Xie:2017erh} attention was given to $N^*(1535)$ production in the  $\Lambda_c^+ \rightarrow \bar{K}^0 \eta p$ reaction. In~\cite{Pavao:2018wdf}, the idea was retaken trying to also find a signature for $N^*(1650)$ production. It was found there that, while in the  $\Lambda_c^+ \rightarrow \bar{K}^0 \pi N$  reaction both resonances could be seen, in the $\Lambda_c^+ \rightarrow \bar{K}^0 \eta p$ reaction only the $N^*(1535)$ was produced.  However, Belle's data showed two distinct peaks in the $\eta p$ channel, corresponding to both $N^*(1535)$ and $N^*(1650)$, indicating the need for a better theoretical understanding.

An important feature overlooked in previous studies~\cite{Xie:2017erh, Pavao:2018wdf} is the role of the external emission mechanism. While internal emission directly produces the $\bar{K}^0 \eta p$ final state at the tree level, external emission does not produce it. Although no $\eta p$ is directly formed in this mechanism, final state interactions can lead to $\eta p$ formation and generate the resonances. Since these resonances arise from meson-baryon rescattering, external emission followed by FSI may even be more relevant for producing the $N^*$ peaks than internal emission.

Motivated by recent experimental observations and theoretical developments, we revisit the decay process $\Lambda_c^+ \rightarrow \bar{K}^0 \eta p$, aiming to deepen our understanding of the underlying hadronization mechanisms and the role of FSI in the formation of baryon resonances. In our analysis, we incorporate both internal and external emission mechanisms, which allow us to study not only the interference between different production topologies but also the dynamical generation of $N^*$ states through strong meson-baryon rescattering.
We first study the decay at the quark level, and through hadronization inserting $q \bar q$ components, we produce two mesons and one baryon in the first step. Later on we consider final state interaction of these components to finally attain the final $\bar{K}^0 \eta p$ state~\cite{Miyahara:2015cja, Xie:2016evi, Dai:2018hqb}.

We employ a coupled-channel chiral unitary approach, where both pseudoscalar-baryon ($PB$) and vector-baryon ($VB$) interactions are included to account for their significant roles in the formation of $N^*(1535)$ and $N^*(1650)$. 
The meson-baryon scattering amplitudes are constructed using effective Lagrangians based on chiral symmetry and hidden local symmetry, and are unitarized through the Bethe-Salpeter equation\cite{Oller:2000fj, Hyodo:2003qa, Roca:2005nm, Doring:2009yv, Nieves:2001wt}. The dynamically generated resonances manifest as poles in the unphysical Riemann sheets of the amplitudes and produce characteristic structures in the invariant mass distributions\cite{Garzon:2014ida}.
This formalism has been successfully applied in related heavy baryon decay studies~\cite{Oset:2016lyh, Sakai:2017hpg, Dong:2021juy}, and provides a reliable framework to investigate the nature and relative contributions of the two resonances in the $\eta p$ invariant mass spectra.

In this work, we aim to:
\begin{enumerate}
\item Provide a theoretical analysis of the $\Lambda_c^+ \rightarrow \bar{K}^0 \eta p$ decay using a coupled-channel chiral unitary framework that includes both $PB$ and $VB$ interactions.
\item Investigate the contributions of dynamically generated $N^*$ resonances, particularly $N^*(1535)$ and $N^*(1650)$, to the $\eta p$ mass spectra.
\item Evaluate the relative importance of internal and external emission mechanisms, and clarify the role of final state interactions in shaping the observed resonance structures.
\item Study the role of the $\Sigma(1620)$ in the $\bar{K}^0 p$ invariant mass distribution.
\item Compare our predictions with available experimental data to test the validity of our model and provide insights into the structure of $N^*$ states.
\end{enumerate}
Through this refined treatment, we aim to understand the interplay between different weak decay mechanisms and final state interactions, and to clarify the origin of the resonance structures observed in the Belle experiment.

This study not only deepens our understanding of charmed baryon decay mechanisms but also supports the continued effort to map the baryon spectrum, with particular emphasis on the role of hadronic molecular components in baryon excitations~\cite{Hyodo:2011ur,Oset:2016lyh,Guo:2017jvc,Meissner:1999vr,Hosaka:2016pey,Roca:2015tea,Oller:1998zr}.

\section{Formalism}
\label{sec_formalism}
\begin{figure}[H]
 \centering
 \includegraphics[width=6cm]{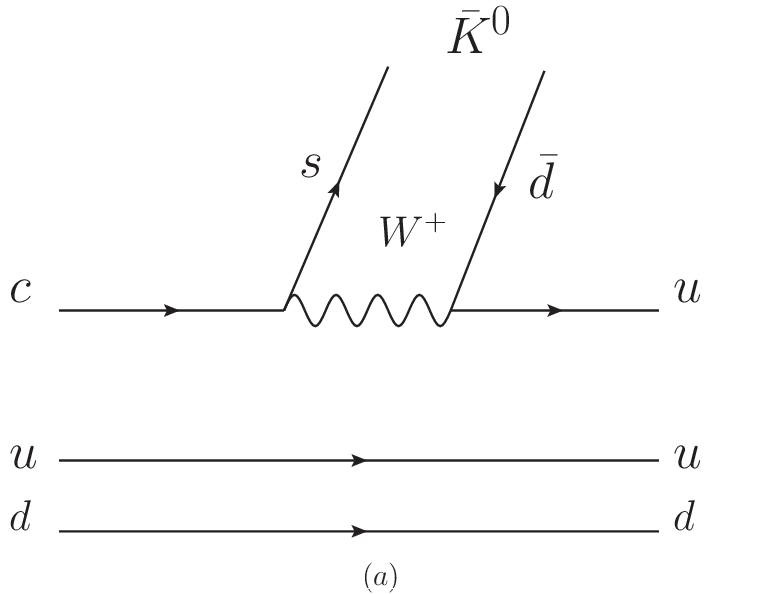}
\qquad\qquad\qquad\qquad \includegraphics[width=6cm]{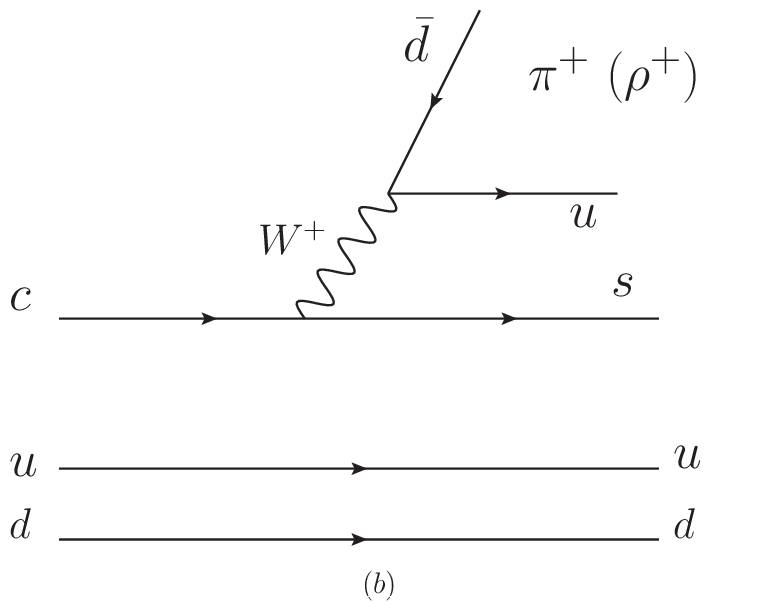}
 \caption{Schematic diagrams for the decay $\Lambda_c^+ \to \bar{K}^0 M B$. (a) internal emission; (b) external emission.}
 \label{fig1}
\end{figure}

In this section, we study the internal and external emission mechanisms following the formalism developed in Ref.~\cite{Xie:2017erh,Pavao:2018wdf}. Both mechanisms are analyzed in detail to understand their contributions to the reaction process.
We study the decay process $\Lambda_c^+ \to \bar{K}^0 M B$, where $M$ and $B$ denote a meson and a baryon, respectively. The main decay mechanisms are illustrated in Fig.~\ref{fig1}.

The hadronization mechanism is modeled from the quark-level transition $c \to s u \bar{d}$. In the internal emission mode, Fig.~\ref{fig1} (a), the $s\bar d$ pair forms the $\bar{K}^0$. The remaining $u u d$ system hadronizes via the insertion of a vacuum $q\bar{q}$ pair, leading to the formation of meson-baryon pairs~\cite{Miyahara:2015cja, Xie:2016evi, Dai:2018hqb}. These pairs then interact through final state interactions, which dynamically generate the $N^*(1535)$ and $N^*(1650)$ resonances. 

In the external emission mode, Fig.~\ref{fig1} (b) the $\bar d u$ pair forms a $\pi^+~(\rho^+)$ and the remaining $sud$ quarks hadronize to give, for instance  $\bar{K}^0 n$. The $\pi^+ \bar{K}^0 n~(\rho^+\bar{K}^0 n)$ state is not the desired final state, but through a $\pi^+n\to \eta p~(\rho^+n\to \eta p)$ transition in the FSI, we obtain the  $\bar{K}^0 \eta p$  final state

\subsection{Primary Weak Decay and Hadronization}
\label{form}

We have two ways to produce  $\bar{K}^0 \eta p$   in the final state: As a first step in internal emission prior to any rescattering, or through rescattering in external emission. We detail the mechanisms below.

\subsubsection{Internal emission}
\label{formalism:in}
\begin{figure}[H]
 \centering
 \includegraphics[width=6cm]{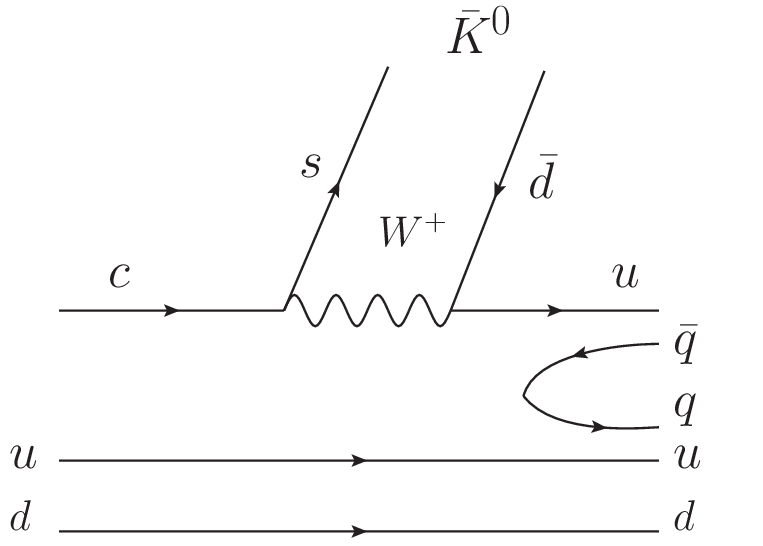}
 \caption{Quark-level diagram for the $\Lambda_c^+ \to \bar{K}^0 M B$ decay.}
 \label{fig:internal_hadronization}
\end{figure}

For internal emission, we follow the same formalism  as in Ref.~\cite{Xie:2017erh,Pavao:2018wdf}, which we describe here for completeness. The diagrams describing hadronization at the quark level are shown in Fig.\ref{fig:internal_hadronization}.
The initial vertex of $\Lambda_c^+ \to \bar{K}^0 M B$ decay is modeled following the approach of Model I in Ref.~\cite{Pavao:2018wdf}. 

The reaction proceeds via a Cabibbo-allowed process involving $W \bar c s$ and $W \bar{d}u$ couplings~\cite{Chau:1982da}, followed by the creation of a light $q \bar{q}$ pair from the vacuum. The $\bar{d} s$ quarks form the $\bar{K}^0$, and the remaining $uud$ quarks with the $\bar{q} q$ hadronize into the meson-baryon final state. The $ud$ pair in $\Lambda_c^+$ acts as a spectator with spin $S=0$ and isospin $I=0$ in the weak process. The $\Lambda_c^+$  state is written as~\cite{Capstick:1986ter,Roberts:2007ni} 
\begin{align}
    \Lambda_c^+= \frac{1}{\sqrt{2}} ~c(ud-du)\chi_\text{MA}
\end{align}
with $\chi_\text{MA}$ the mixed antisymmetric spin wave function.

At the quark level, the final state can be written as
\begin{align}
 \bar{K}^0 ~u(\bar{u}u + \bar{d}d + \bar{s}s) \frac{1}{\sqrt{2}} (ud - du) = \bar{K}^0 ~\sum_i P_{1i} q_i \frac{1}{\sqrt{2}} (ud - du),
\end{align}
where $P_{ij} = q_i \bar{q}_j$ and $q_1 = u$, $q_2 = d$, $q_3 = s$. The matrix $P$ corresponds to the meson matrix in flavor space~\cite{Pavao:2017cpt,Bramon:1992kr}:
\begin{align}\label{pmatrix}
 P = 
 \begin{pmatrix}
  \frac{\pi^0}{\sqrt{2}} + \frac{\eta}{\sqrt{3}} + \frac{\eta'}{\sqrt{6}} & \pi^+ & K^+ \\
  \pi^- & -\frac{\pi^0}{\sqrt{2}} + \frac{\eta}{\sqrt{3}} + \frac{\eta'}{\sqrt{6}} & K^0 \\
  K^- & \bar{K}^0 & -\frac{\eta}{\sqrt{3}} + \sqrt{\frac{2}{3}} \eta'
 \end{pmatrix},
\end{align}
where the $\eta-\eta'$ mixing of Ref.~\cite{Bramon:1992kr} is used. The $\eta'$ does not play a role in the  interaction and we do not consider it here.
Therefore, the final hadronic state can be expanded as
\begin{align}
 &\sum_i P_{1i} q_i \frac{1}{\sqrt{2}} (ud - du) \notag \\
 &= \left(\frac{\pi^0}{\sqrt{2}} + \frac{\eta}{\sqrt{3}}\right) u \frac{1}{\sqrt{2}} (ud - du) + \pi^+ d \frac{1}{\sqrt{2}} (ud - du) + K^+ s \frac{1}{\sqrt{2}} (ud - du).    
\end{align}

Using the mixed antisymmetric quark configurations of the baryons~\cite{Miyahara:2016yyh,Pavao:2017cpt},
\begin{align}
 p &= \frac{u(ud - du)}{\sqrt{2}}, \\
 n &= \frac{d(ud - du)}{\sqrt{2}}, \\
 \Lambda &= \frac{u(ds - sd) + d(su - us) - 2 s(ud - du)}{2\sqrt{3}},
\end{align}
the meson-baryon final state excluding $\bar{K}^0$ is expressed as
\begin{align}
 |PB\rangle~\chi_\text{MA} =& \Bigg(\frac{1}{\sqrt{2}} |\pi^0 p\rangle + \frac{1}{\sqrt{3}} |\eta p\rangle + |\pi^+ n\rangle - \sqrt{\frac{2}{3}} |K^+ \Lambda\rangle \Bigg)~\chi_\text{MA}
\end{align}
and considering the overlap with the baron octet states, $\phi=\frac{1}{\sqrt{2}} (\phi_\text{MS}\chi_\text{MS}+\phi_\text{MA}\chi_\text{MA})$, then we have~\footnote{This factor, \(1/\sqrt{2}\), was omitted in  Refs.~\cite{Pavao:2018wdf,Xie:2017erh}, with no consequence in the results that depend on an arbitrary normalization factor.}

\begin{align}
 |PB\rangle =& \frac{1}{2} |\pi^0 p\rangle + \frac{1}{\sqrt{6}} |\eta p\rangle + \frac{1}{\sqrt{2}} |\pi^+ n\rangle - \sqrt{\frac{1}{3}} |K^+ \Lambda\rangle, 
 \label{eq:pb_channels}
\end{align}
The $\pi N$ state can be  written in the isospin basis (with the convention $|\pi^+\rangle = -|I=1, I_z=1\rangle$), and then we have
\begin{align}
 |PB\rangle  =& - {\frac{\sqrt{3}}{2}} |\pi N \rangle + \frac{1}{\sqrt{6}} |\eta p\rangle - \sqrt{\frac{1}{3}} |K^+ \Lambda\rangle,
 \label{eq:pb_channels_isospin}
\end{align}
which reflects the $I=1/2$ of the $u$ quark and the $I=0$ of the $ud$ pair in a $u(ud-du)$ of the diagram of Fig.~\ref{fig1} (a).

Similarly, the vector meson matrix $V$  is
\begin{align}
 V =
 \begin{pmatrix}
  \frac{\rho^0}{\sqrt{2}} + \frac{\omega}{\sqrt{2}} & \rho^+ & K^{*+} \\
  \rho^- & -\frac{\rho^0}{\sqrt{2}} + \frac{\omega}{\sqrt{2}} & K^{*0} \\
  K^{*-} & \bar{K}^{*0} & \phi
 \end{pmatrix},
\end{align}
The same hadronization mechanism discussed above gives rise to the vector meson-baryon state
\begin{align}
 |VB\rangle = - {\frac{\sqrt{3}}{2}}~ |\rho N (I=1/2)\rangle,
 \label{eq:vb_channels}
\end{align}
and following Ref.~\cite{Pavao:2018wdf}, we keep the $\rho N$ component, found the most relevant ingredient to generate the $N^*(1650)$   state in~\cite{Garzon:2014ida}.

Combining the pseudoscalar meson-baryon and vector meson-baryon states, the final hadronic state (excluding $\bar{K}^0$) is
\begin{align}
  \sum_{MB} h_{MB} |MB\rangle = & - {\frac{\sqrt{3}}{2}}~ |\pi N (I=1/2)\rangle + \frac{1}{\sqrt{6}}~ |\eta N\rangle - \sqrt{\frac{1}{3}}~ |K^+ \Lambda\rangle  - {\frac{\sqrt{3}}{2}}~ |\rho N (I=1/2)\rangle.
 \label{eq:pb_final}
\end{align}
\noindent We call \( h_{MB} \)   the   weights of each meson-baryon channel of Eq.~(\ref{eq:pb_final}), which are summarized in Table~\ref{tab:weights}.

\begin{table}[H]
 \centering
\setlength{\tabcolsep}{28pt}
\begin{tabular}{c|cccc}
  Channel & $\pi N $ & $\eta N$ & $K \Lambda$ & $\rho N  $ \\
  \hline
  $h_{MB}$ & $-{\sqrt{3}}/{2}$ & ${1}/{\sqrt{6}}$ & $-{{1}/\sqrt{3}}$ & $-{\sqrt{3}}/{2}$ \\
 \end{tabular}
 \caption{Relative weights $h_{MB}$ of meson-baryon channels in the primary decay vertex.}
 \label{tab:weights}
\end{table}
However, as pointed out in Ref.~\cite{Pavao:2018wdf} and also adopted in this work, an additional factor arises from the hadronization into either \( VB \) or \( PB \) channels, as discussed in Refs.~\cite{Liang:2016ydj,Pavao:2017cpt}. This factor originates from the explicit implementation of the \( ^3P_0 \) model in the hadronization process and is given by  
\begin{align}
    f_{PB} &= \frac{1}{4\pi} \cdot \frac{1}{2}, \\
    f_{VB} &= \frac{1}{4\pi} \cdot \frac{1}{2\sqrt{3}}.
\end{align}
\noindent We take this into account by changing $h_{MB}$ to $h'_{MB}$ where 
$$ h'_{\rho N}=  h_{\pi N}/\sqrt{3}. $$

Consequently, we have
\begin{align}
    h'_{\pi N} = -\frac{\sqrt{3}}{2},\qquad
    h'_{\eta N} = \frac{1}{\sqrt{6}},    \qquad 
    h'_{K\Lambda} = -\sqrt{\frac{1}{3}},\qquad
    h'_{\rho N} = -\frac{1}{2}.
\end{align}

Combining these with the additional  $\bar{K}^0$ from Fig.\ref{fig:internal_hadronization}, we obtain the hadronic final state arising from the internal emission
\begin{align}
\Lambda_c^+ \longrightarrow  \bar K^0~ \sum_{MB} h'_{MB} |MB\rangle =   & \bar K^0~\Bigg\{h'_{\pi N}~ | \pi N \rangle + h'_{\eta N}~ |\eta N\rangle +h'_{K\Lambda}~ |K^+ \Lambda\rangle + h'_{\rho N}~ |\rho N  \rangle 
 \Bigg\}.
 \label{eq:mb_final}
\end{align}

\subsubsection{External Emission}
\label{formalism:ex}

\begin{figure}[H]
\centering
\includegraphics[scale=0.48]{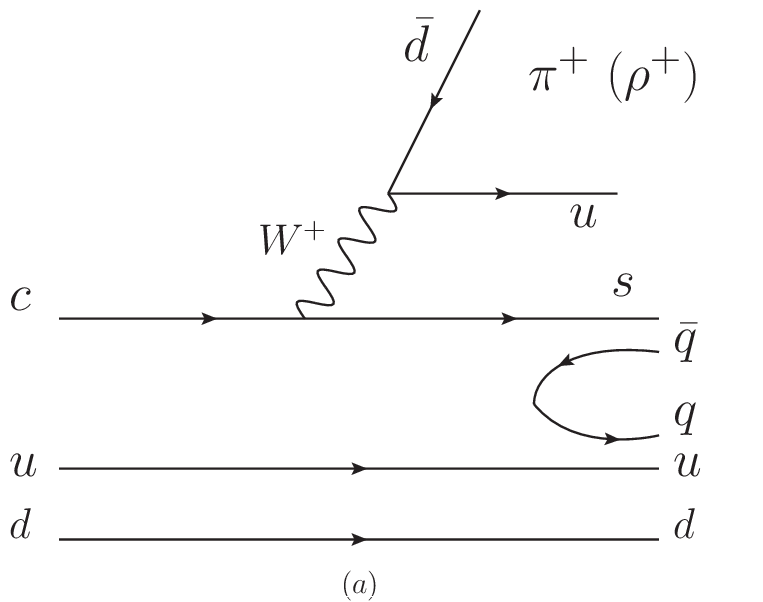} \qquad\qquad\qquad\qquad
\includegraphics[scale=0.48]{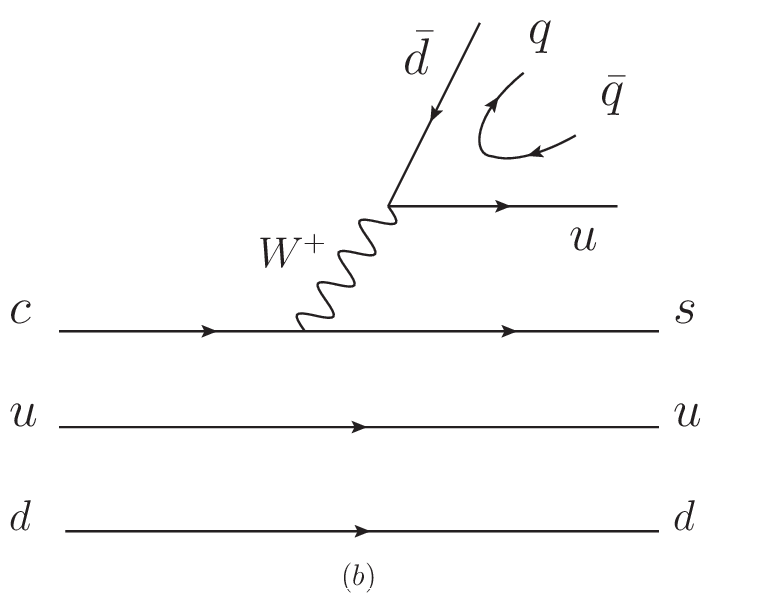}
\caption{Hadronization in external emission: (a) of the $su~(sd)$ pair; (b) of the $u\bar{d}$ pair.}
\label{fig:external_hadronization}
\end{figure}

The external emission mechanism, which is color-favored, is illustrated at the quark level in Fig.\ref{fig:external_hadronization}, where we show two possible hadronizations  of quark pairs leading finally to two mesons and one baryon. 
In Fig.~\ref{fig:external_hadronization} (a), the $\bar{d} u$ quarks form the $\pi^+$ or $\rho^+$, and the remaining $sud$ quarks hadronize with the $\bar{q} q$  into the meson-baryon final state. The \(ud\) pair in \(\Lambda_c^+\) acts as a spectator in the weak decay process of Figs.~\ref{fig:external_hadronization} (a) and \ref{fig:external_hadronization} (b). In Fig.~\ref{fig:external_hadronization} (a), together with the \(\pi^+\), the hadronization of the \(s {u}~ (sd)\) quark pair proceeds as

\begin{align}
\pi^+  \sum_i s \bar{q}_i q_i(ud-du)\frac{1}{\sqrt{2}}\chi_\text{MA},
\end{align}
where $q_i$ denotes the light quark flavors $u$, $d$, $s$. This can be expressed in terms of the pseudoscalar meson fields $P$ matrix of Eq.~(\ref{pmatrix}) as
\begin{align}
\pi^+\frac{1}{\sqrt{2}} \Bigg[P_{31}u(ud-du)+P_{32}d(ud-du)+P_{33}s(ud-du)\Bigg]\chi_\text{MA}.
\end{align}
By considering the overlap with the octet of baryons, 
\(\phi = \frac{1}{\sqrt{2}} (\phi_\text{MS}\chi_\text{MS} + \phi_\text{MA}\chi_\text{MA})\), we obtain
\begin{align}
    \Lambda_c^+ \longrightarrow \pi^+ \left( \frac{1}{\sqrt{2}} |K^- p\rangle + \frac{1}{\sqrt{2}} |\bar K^0 n\rangle + \frac{1}{3} |\eta \Lambda\rangle \right).
    \label{eq:pb_channels_ex}
\end{align}

\noindent
None of these states is the \( \bar K^0 \eta p \) final state, but we can obtain it via the \( \pi^+ n \to \eta p \)  interaction.
For the transition \( \pi^+ n \to \eta p \), we need to express the initial state in terms of isospin eigenstates for the process \( \pi N \to \eta N \):
\begin{align}\label{ex_FS}
    |\pi^+ n\rangle = -\left( \sqrt{\frac{1}{3}}\, |3/2,\,1/2\rangle + \sqrt{\frac{2}{3}}\, |1/2,\,1/2\rangle \right),
\end{align}
\noindent
where the negative sign arises from the isospin convention for the \( \pi^+ \) state. Consequently, the transition amplitude becomes
\begin{align}\label{ex_t}
    t_{\pi^+ n,\, \eta p} = -\sqrt{\frac{2}{3}}\, t_{\pi N,\, \eta N}.
\end{align}

Let us now turn the attention to the mechanism of Fig.~\ref{fig:external_hadronization} (b). We have a final $\Lambda$ baryon state and with the hadronization of the $u \bar d$ pair we would get
\begin{align}\label{eq:vanishFS}
    u \bar d\Lambda &\longrightarrow \sum_i u\bar q_iq_i\bar d\Lambda=\sum_i P_{1i}P_{i2}\Lambda=(P^2)_{12}\Lambda \\\nonumber
    & = \Bigg\{(\frac{\pi^0}{\sqrt{2}} + \frac{\eta}{\sqrt{3}}) \pi^+ + \pi^+(-\frac{\pi^0}{\sqrt{2}} + \frac{\eta}{\sqrt{3}}) +K^+\bar K^0 \Bigg\}~\Lambda
\end{align}
and while from the channel $\eta\pi^+\Lambda$ we could have  $\pi^+\Lambda\to \bar K^0p$, and from the $K^+\bar K^0 \Lambda$ we could have $K^+\Lambda\to \eta p$ and have the desired final state, one must be cautious. Indeed, one has to look in detail into the $WPP$ vertex by using effective chiral Lagriangians~\cite{Gasser:1983yg,Scherer:2002tk,Ren:2015bsa},

\begin{align}
    \mathcal{L}=\frac{f^2}{4}\left\langle D_\mu U\left(D^\mu U\right)^{\dagger}\right\rangle+\frac{f^2}{4}\left\langle\chi U^{\dagger}+U \chi^{\dagger}\right\rangle,
\end{align}
where
\begin{align}
&D_\mu U=\partial_\mu U-i r_\mu U+i U l_\mu,\nonumber\\
&U= \textrm{exp}(i\frac{\sqrt{2}P}{f}),
\end{align}
with $r_\mu,~l_\mu$
 the right-handed, left-handed weak interaction  currents
 \begin{align}
&r_\mu=v_\mu+a_\mu  = e Q A_\mu + \cdot\cdot\cdot, \nonumber\\
&l_\mu=v_\mu-a_\mu=e Q A_\mu+\frac{e}{\sqrt{2}\sin{\theta}_W}\left(W_\mu^{\dagger} T_{+}+W_\mu T_{-}\right),
\end{align}
where
\begin{align}
T_{+}=\left(\begin{array}{ccc}
0 & V_{u d} & V_{u s} \\
0 & 0 & 0 \\
0 & 0 & 0
\end{array}\right), \quad T_{-}=\left(\begin{array}{ccc}
0 & 0 & 0 \\
V_{u d} & 0 & 0 \\
V_{u s} & 0 & 0
\end{array}\right),
\end{align}
with $V_{u d},~V_{u s}$, matrix elements of the Cabibbo–Kobayashi–Maskawa matrix, 
where, $v_\mu,~a_\mu,~A_\mu$, and $W_\mu$ are the vector current, axial current, electromagnetic field, and $W$ field,  $q$ the electron charge and $Q$ the matrix of the  charge of quarks $u,~d,~s$. We are only concerned about the weak part of $l_\mu$.
After expansion of the fields, one finds 
for the $WPP'$ vertex of Fig. \ref{fig:external_hadronization} (b)

\begin{align}\label{eq:L}
\mathcal{L} = -i \frac{1}{2} \frac{e}{\sqrt{2} \sin \theta_W} W_\mu^+ \langle [P, \partial_\mu P] T_+ \rangle,
\end{align}
which determines  the $WPP'$ vertex. 

What matters for us here is 
the structure of the  term $ \langle [P,~\partial_\mu P] W^{\mu}\rangle  $. Indeed, the terms $\eta \pi^+$ and $\pi^+\eta$ of Eq.~(\ref{eq:vanishFS}) cancel rather than sum because of the  $[P, \partial_\mu P]$  commutator. Then we have the $\bar K^0K^+$ term.
The term $K^+\bar{K}^0$ could contribute through $K^+\Lambda \to \eta p$, but the structure of Eq.~(\ref{eq:L}), $K^+\partial_\mu \bar{K}^0 - \bar{K}^0 \partial_\mu K^+$, with the dominant $\mu=0$ component, makes this term vanish in an average through the calculation of the difference of the $\bar{K}^0$ and $K^+$ energies. Therefore, this external emission mechanism of fig.~\ref{fig:external_hadronization} (b) does not contribute to the final $\bar{K}^0 \eta p$ state.~\footnote{In the work of Ref.~\cite{Xie:2017erh} a diagram like our Fig.~\ref{fig:external_hadronization} (b) is considered with $q\bar q$ being $s\bar s$, but considered as a pentaquark component for the $N^*(1535)$ wave function. There is no objection to this assumption, but from the strict molecular picture for the  $N^*(1535)$,  $N^*(1650)$ as made of meson baryon components, the topology of Fig.~\ref{fig:external_hadronization} (b) does not lead to the desired final sate, as we have shown.}

\subsection{Final State Interaction and Decay Amplitude}
\begin{figure}[H]
  \begin{center}
  \includegraphics[scale=0.4]{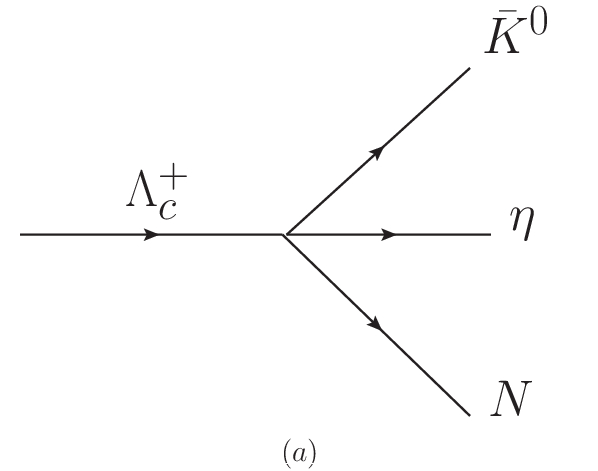}
  \includegraphics[scale=0.4]{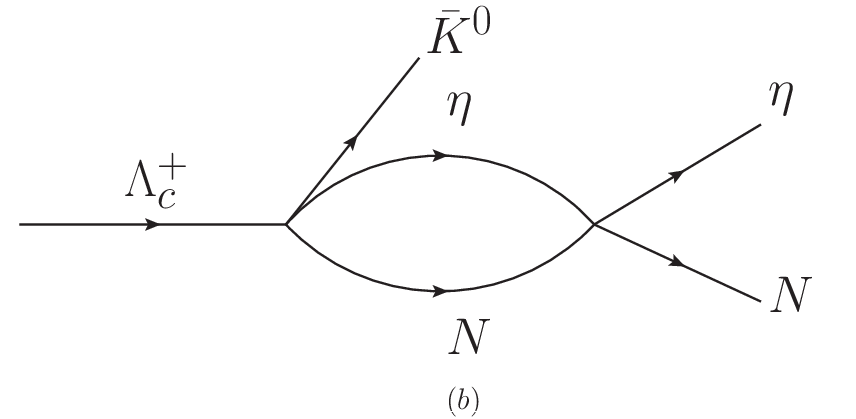}
  \includegraphics[scale=0.4]{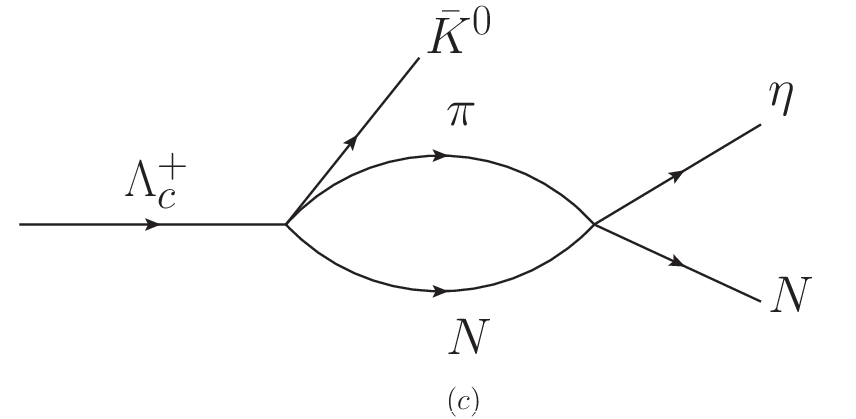}
  \includegraphics[scale=0.4]{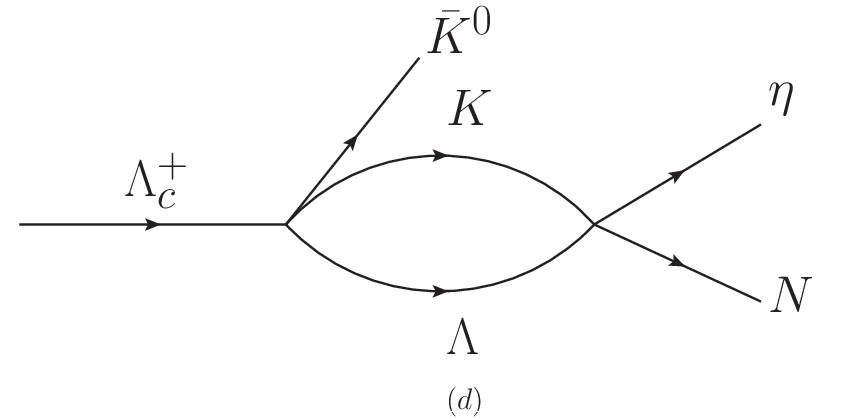}
  \includegraphics[scale=0.4]{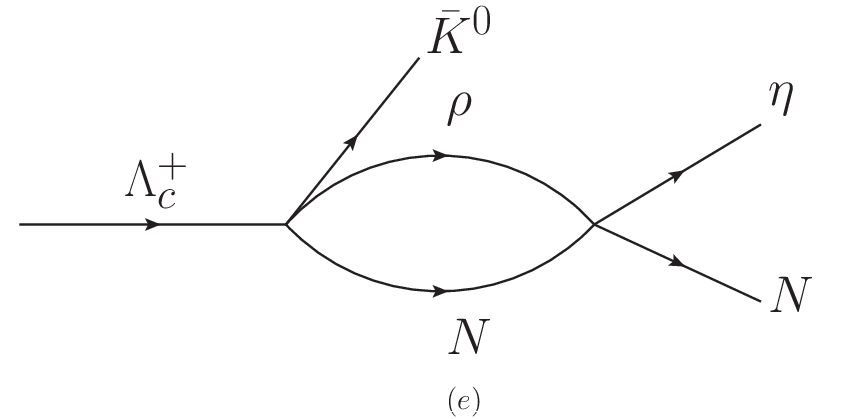}
  \end{center}
  \vspace{-0.5cm}
  \caption{Diagrams showing the production of $\bar K^0 \eta p $ from internal emission and external emission: (a) tree level; (b) $ \eta N$ rescattering; (c) $\pi  N$ rescattering; (d) $K\Lambda$ rescattering; (e) $\rho  N$ rescattering.}
  \label{fig:FSI_FEYMANN}
  \end{figure}

Summing the two mechanisms, including the $\bar K^0$ from Fig.~\ref{fig:internal_hadronization}   and $\pi^+~(\rho^+)$ from Fig.~\ref{fig:external_hadronization} (a), we observe a tree-level production of $\bar K^0 \eta p$, with the other terms potentially leading to this final state via rescattering, as shown in Fig.~\ref{fig:FSI_FEYMANN}.

The decay amplitude includes the primary production and subsequent meson-baryon rescattering effects. The rescattering involves coupled channels $MB = \pi N (I=1/2), \eta N, K \Lambda, \rho N (I=1/2)$, described by the scattering matrix $T_{MB \to MB}$  with a parametrized amplitudes as Breit-Wigner,
 \begin{align}\label{ampTW}
t_{i,j} = \frac{g_ig_j}{M_\text{inv}-M_R+i\frac{\Gamma_R}{2}}.
 \end{align}
The couplings to $N^*(1535)$ and $N^*(1650)$ are given in Table~III of Ref.~\cite{Garzon:2014ida}, which we show in Table~\ref{tab:couplings} here.
\begin{table}[H]
\centering
\caption{Couplings to $N^*(1535)$  and $N^*(1650)$ from Ref.~\cite{Garzon:2014ida}}
\label{tab:results_decuplet550}
\setlength{\tabcolsep}{48pt}
\begin{tabular}{ccc}
\hline \hline
Coupling & $N^*(1535)$ & $N^*(1650)$  \\
\hline
$g_{\pi N}$      & $1.03 + i\,0.21 $ & $1.37 + i\,0.54$ \\
$g_{\eta N}$     & $1.40 + i\,0.78$                              & $1.08 - i\,0.60$ \\
$g_{K \Lambda}$  & $1.71 + i\,0.48$                              & $0.10 - i\,0.68$ \\
$g_{\rho N}$     & $2.96 + i\,0.11$                              & $0.94 + i\,1.51$ \\
\hline\hline
\end{tabular}
\label{tab:couplings}
\end{table}
\noindent
Due to the proximity of the \( \eta N \) threshold, the width \( \Gamma_R^{N^*(1535)} \) should be taken as energy-dependent. We take both  $\Gamma_{\pi N}$, $\Gamma_{\eta N}$  as  energy-dependent given by
\begin{align}
    \Gamma_R^{N^*(1535)}(M_\text{inv}) = \Gamma_{\pi N}(M_\text{inv}) + \Gamma_{\eta N}(M_\text{inv}).
\end{align}
\noindent
In contrast, for the \( N^*(1650) \), we take a constant width \( \Gamma_R^{N^*(1650)} = 125~\text{MeV} \), as quoted in Ref.~\cite{ParticleDataGroup:2024cfk}.

The partial widths \( \Gamma_{\pi N}(M_\text{inv}) \) and \( \Gamma_{\eta N}(M_\text{inv}) \) are given by:
\begin{align}
    \Gamma_{\pi N}(M_\text{inv}) = \frac{1}{2\pi} \frac{M_N}{M_\text{inv}}|g_{\pi N}|^2q_\pi(M_\text{inv}),\\\nonumber
    \Gamma_{\eta N}(M_\text{inv}) = \frac{1}{2\pi} \frac{M_N}{M_\text{inv}}|g_{\eta N}|^2q_\eta(M_\text{inv}),    
\end{align}
with 
\begin{align}
     q_{i} = \frac{\lambda^{1/2}(M_\text{inv}^2, m_{M_i}^2, M_N^2)}{2M_\text{inv}}, \qquad {i} = \pi ~\text{or}~ \eta.  
\end{align}
where $\lambda$ is the Källén function.
We adopt the model of Ref.~\cite{Garzon:2014ida,Pavao:2018wdf} for the meson-baryon interaction, where the loop functions $G_{MB}$ are evaluated with dimensional regularization and given by 
\begin{align}\label{eq_loop_anal}
 G_{MB}(\sqrt{s},m_M,M_B)=&\frac{2M_B}{16\pi^2}\{a_{MB}(\mu)+\ln\frac{M_B^2}{\mu^2}+\frac{m_M^2-M_B^2+s}{2s}\ln\frac{m_M^2}{M_B^2}\notag\\
 &\hspace{5mm}+\frac{q_{MB}}{\sqrt{s}}\left[\ln(s-M_B^2+m_M^2+2q_{MB}\sqrt{s})\right.\notag\\
 &\hspace{1.5cm}+\ln(s+M_B^2-m_M^2+2q_{MB}\sqrt{s})\notag\\
 &\hspace{1.5cm}-\ln(-s+M_B^2-m_M^2+2q_{MB}\sqrt{s})\notag\\
 &\hspace{1.5cm}\left.-\ln(-s-M_B^2+m_M^2+2q_{MB}\sqrt{s})\right]\},
\end{align}
where $\sqrt{s}$ is the invariant mass of the $MB$ pair.
The subtraction constants $a_\mu$ are evaluated using the dimensional regularization scheme, where $\mu$ is the regularization scale. Following Ref.~\cite{Garzon:2014ida}, we take $\mu = M_B$ for each channel, with $M_B$ the mass of the baryon and $m_M$ the mass of the meson involved in the loop function. The corresponding values of the subtraction constants are taken from Table II of Ref.~\cite{Garzon:2014ida} and given by
\begin{align}
a_{N \pi} = -1.203, \qquad
a_{N \eta} = -2.208, \qquad
a_{\Lambda K} = -1.985, \qquad
a_{N \rho} = -0.493.
\end{align}
And $q_{MB}$ is the momentum of the meson in the center-of-mass (CM) frame of the meson-baryon system:
\begin{align}
 q_{MB} = \frac{\lambda^{1/2}(s, m_M^2, M_B^2)}{2\sqrt{s}}.
\end{align}

The decay amplitude arising from internal emission can the be written as
\begin{align}\label{inamp}
    t^{\text{(int)}} = \mathcal{C} \beta \Bigg(
        &h'_{\eta N} + h'_{\eta N} G_{\eta N}(M_\text{inv})\, t_{\eta N \to \eta N}(M_\text{inv}) + h'_{\pi N} G_{\pi N}(M_\text{inv})\, t_{\pi N \to \eta N}(M_\text{inv}) \nonumber\\
        &+ h'_{K\Lambda} G_{K\Lambda}(M_\text{inv})\, t_{K\Lambda \to \eta N}(M_\text{inv}) 
        + h'_{\rho N} G_{\rho N}(M_\text{inv})\, t_{\rho N \to \eta N}(M_\text{inv})
    \Bigg),
\end{align}
with $M_\text{inv}$ the $\eta p$ invariant mass,
where the first term corresponds to the tree-level production, while the remaining terms account for the strong final-state interactions, which dynamically generate the \( N^*(1535) \) and \( N^*(1650) \) resonances.

The amplitude arising from external emission can be written from Eqs.~(\ref{ex_FS})~(\ref{ex_t})as
\begin{align}\label{examp}
 t^\text{(ext)} = -\mathcal{C}  \sqrt{\frac{1}{3}} \left(  
 G_{\pi N}(M_\text{inv})\, t_{\pi N \to \eta N}(M_\text{inv}) + 
 G_{\rho N}(M_\text{inv})\, t_{\rho N \to \eta N}(M_\text{inv}) 
 \right),
 \end{align}
where we have also included the  $\rho N$  channel.
The parameter $\mathcal{C} $ is an overall weak coupling constant in units of MeV$^{-1}$, with $\beta\sim\frac{1}{N_c}=\frac{1}{3}$ being the corresponding color reduction factor of internal versus external emission. 
With the sum of $t^\text{(ext)}$ and $t^\text{(int)}$, the decay amplitude of the $\Lambda_c^+ \rightarrow \bar{K}^0 \eta p$ process from the diagrams in Figs.~\ref{fig:FSI_FEYMANN}  is given by
\begin{align}
 t_{\Lambda_c^+ \rightarrow \bar{K}^0 \eta p} =& ~t^\text{(int)} + t^\text{(ext)}.
 \label{eq_amp_final}
\end{align}
Note that we take the same $G-$functions in Eqs.~(\ref{inamp})~(\ref{examp}) as those used in the \( PB \) amplitudes \( t_{MB,M'B'} \) from Ref.~\cite{Garzon:2014ida}. 
For the \( VB \) sector, in  the \( \rho N \) channel, we employ the smeared loop function \( \tilde{G}_{\rho N}(\sqrt{s}) \), which is obtained by folding the $G-$function with the spectral function of the \( \rho \) meson:
\begin{align}
 \tilde{G}_{\rho N}(\sqrt{s}) &= \frac{1}{N} \int_{(m_\rho - 2\Gamma_\rho)^2}^{(m_\rho + 2\Gamma_\rho)^2} \, d\tilde{m}^2 \left(-\frac{1}{\pi}\right) \times \text{Im} \left[\frac{1}{\tilde{m}^2 - m_\rho^2 + i~m_\rho \Gamma_\rho} \right] G_{\rho N}(\sqrt{s}, \tilde{m}, M_N),
\end{align}
where
\begin{align}
 N &= \int_{(m_\rho - 2\Gamma_\rho)^2}^{(m_\rho + 2\Gamma_\rho)^2} \, d\tilde{m}^2 \left(-\frac{1}{\pi}\right) \text{Im} \left[\frac{1}{\tilde{m}^2 - m_\rho^2 + i~m_\rho \Gamma_\rho} \right].
\end{align}

\subsection{Differential Decay Width}

Since the $\Lambda_c^+$ decay amplitude depends only on $M_\text{inv}(\eta p)$, the $\eta p$ mass distribution can be obtained as
\begin{align}\label{GeN}
 \frac{d\Gamma_{\Lambda_c^+ \rightarrow \bar{K}^0 \eta p}}{dM_\text{inv}(\eta p)} = \frac{1}{(2\pi)^3} \frac{M_N}{M_{\Lambda_c^+}} {p}_{\bar{K}^0}{\tilde{p}}_\eta  |t_{\Lambda_c^+ \rightarrow \bar{K}^0 \eta p}|^2,
\end{align}
where $|{p}_{\bar{K}^0}|$ and $|{\tilde{p}}_\eta |$ are the momenta of $\bar{K}^0$ in  $\Lambda_c^+$ rest frame and of  $\eta$ in the $\eta p$ CM frame, respectively:
\begin{align}
 {p}_{\bar{K}^0} &= \frac{\lambda^{1/2}(M_{\Lambda_c^+}^2, m_{\bar{K}^0}^2, M_\text{inv}(\eta p)^2)}{2M_{\Lambda_c^+}}, \\
 {\tilde{p}}_\eta &= \frac{\lambda^{1/2}(M_\text{inv}(\eta p)^2, m_\eta^2, M_N^2)}{2M_\text{inv}(\eta p)}.
\end{align}

To calculate the other mass distributions (we label the particles as $\eta$ (1), $p$ (2), $\bar K^0$ (3)), we use the standard PDG formula \cite{ParticleDataGroup:2024cfk}:
\begin{equation}\label{eq:Gamm}
  \dfrac{d^2 \Gamma}{dM_{12}\; dM_{23}} = \dfrac{1}{(2\pi)^3}\; \dfrac{1}{32\, M^3_{\Lambda_c^+}}\; 2M_{\Lambda_c^+}2M_N  |t|^2 2M_{12} 2M_{23}.
\end{equation}
Our matrix $t$ is spin independent and $|t|^2$ already accounts for the sum and average over the final and initial spins of the baryons. 
By applying the limiting cases of the PDG prescription~\cite{ParticleDataGroup:2024cfk}, one recovers the same result as in Eq.~(\ref{GeN}). The differential decay width \( d\Gamma / dM_{12} \) can be obtained by integrating over \( M_{23} \). 

Furthermore, by performing cyclic permutations of the kinematic variables, one can derive the distributions \( d\Gamma / dM_{13} \) and \( d\Gamma / dM_{23} \), which can be directly compared with the experimental data from Ref.~\cite{Belle:2022pwd}.

\section{Results and Discussion} \label{sec:results_discussion}

In this section, we present and analyze  the $\eta p$, $\bar{K}^0 p$, and $\bar{K}^0 \eta$ mass distributions in the decay $\Lambda_c^+ \to \bar{K}^0 \eta p $, focusing on the role of the $N^*(1535)$, $N^*(1650)$ and $\Sigma(1620)$ resonances. These states are dynamically generated from meson-baryon final state interactions within the chiral unitary approach. 

\begin{figure}[H]
  \centering
 \includegraphics[width=0.33\textwidth]{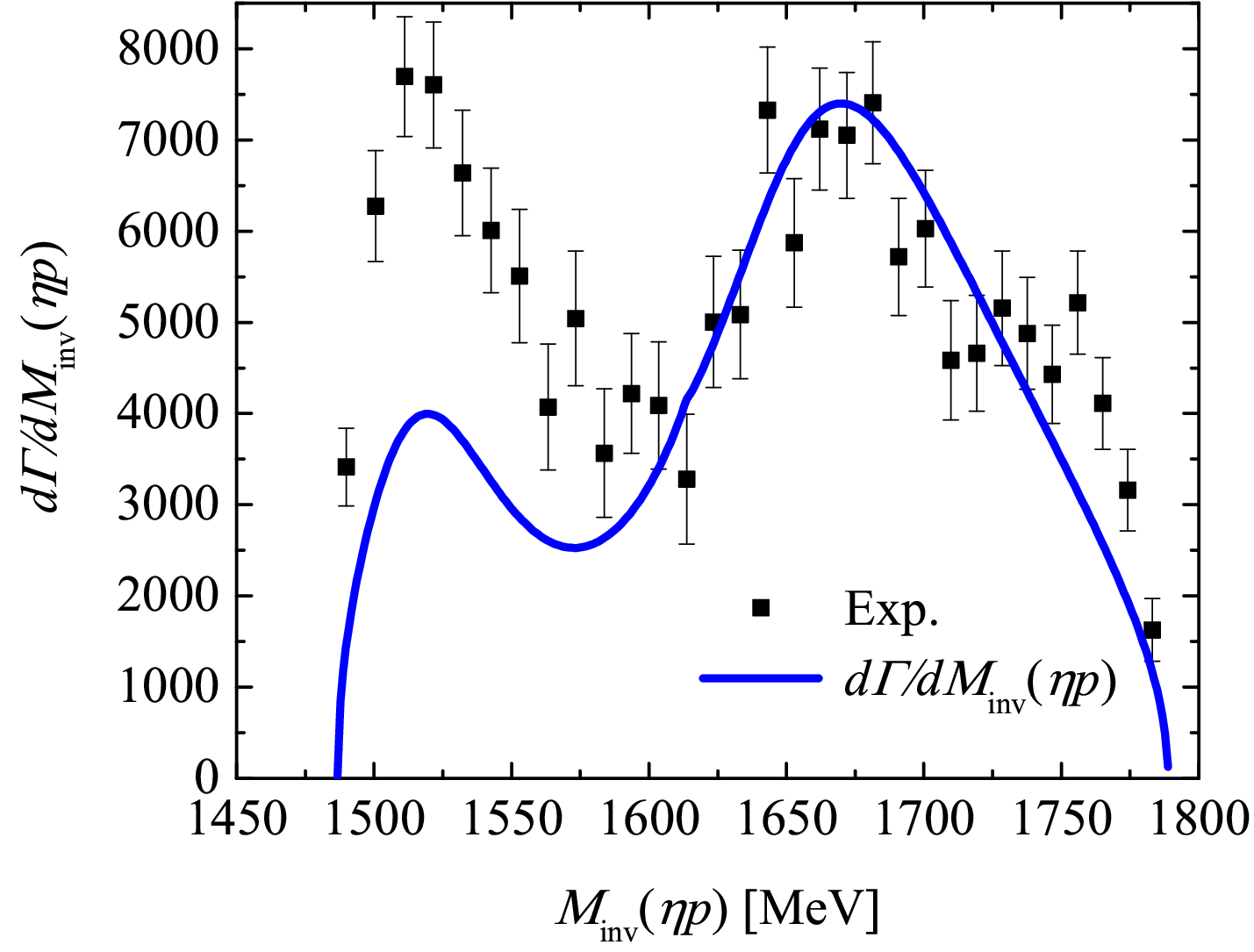}
 \includegraphics[width=0.33\textwidth]{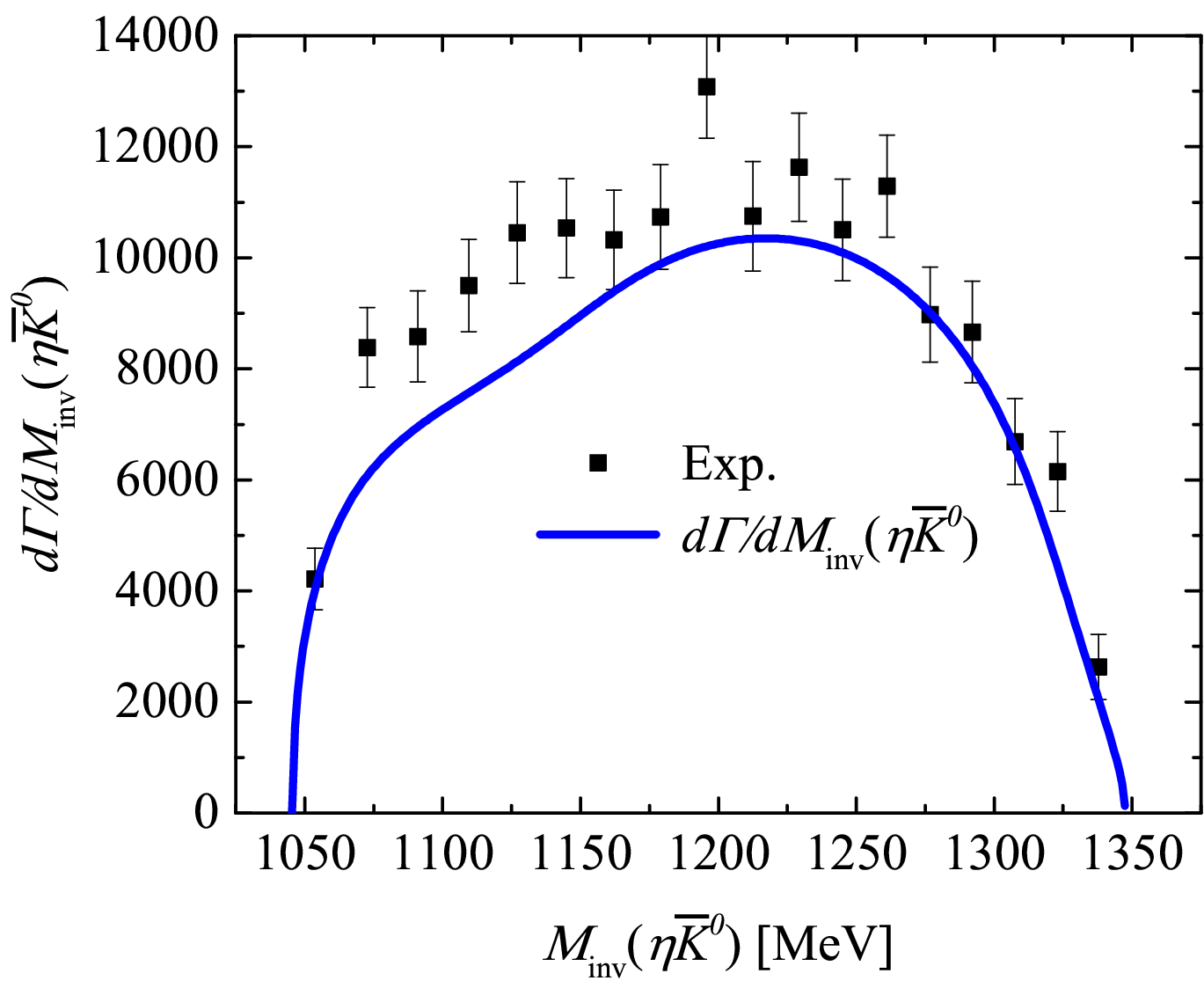}
 \includegraphics[width=0.33\textwidth]{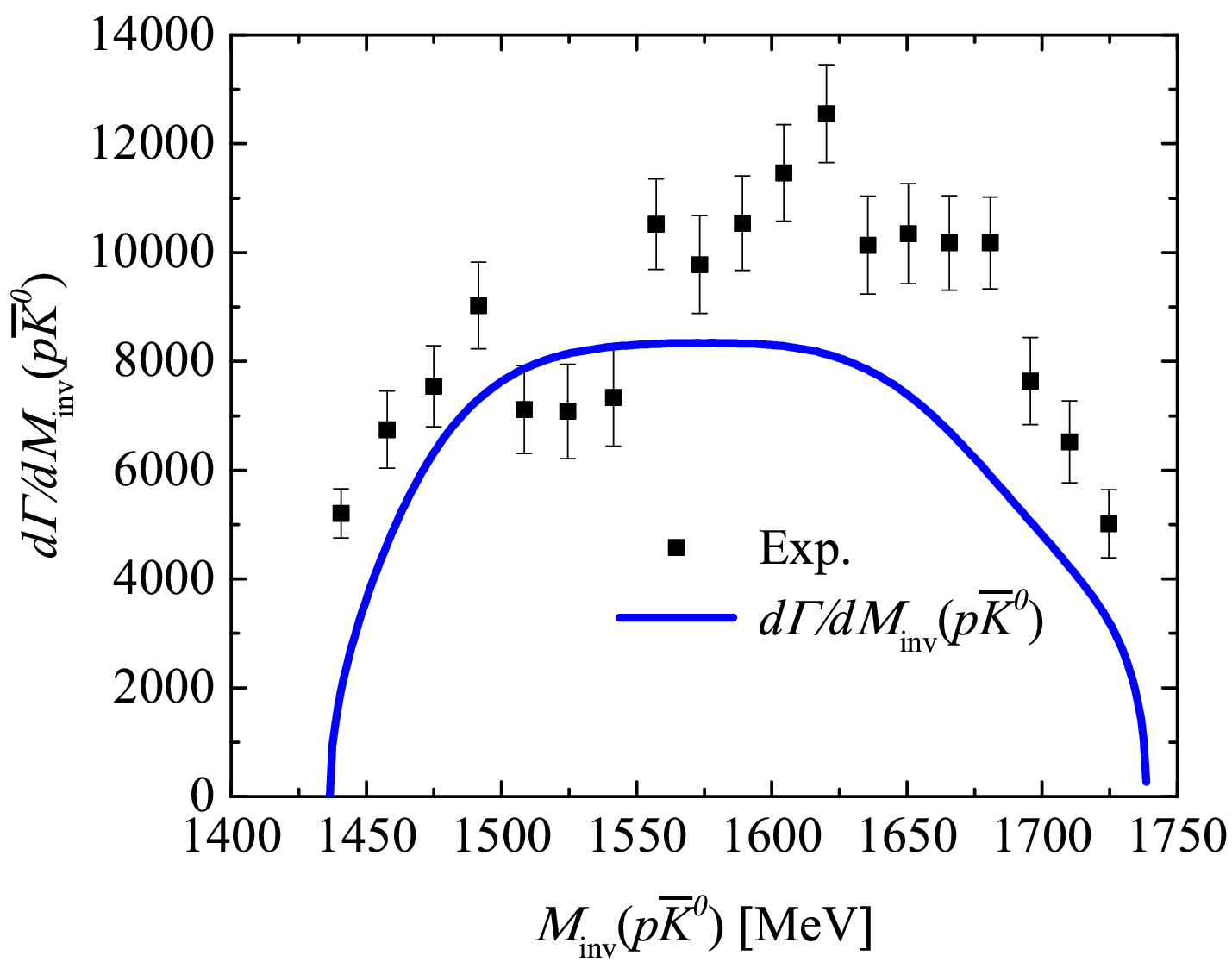}
  \caption{Invariant mass distributions of $\eta p$ (left), $\eta \bar{K}^0$ (middle), and $p \bar{K}^0$ (right) with $M_{N^*(1535)} = 1525$ MeV and $M_{N^*(1650)} = 1650$ MeV fixed. }
  \label{res:results}
\end{figure}

Before performing any tuning of the parameters of the theory, we first present the results obtained with a normalization factor $\mathcal{C}$ and a fixed value $\beta = 1/N_c = 1/3$~\footnote{The binning of the experiments for the mass distributions are different and we have taken that into account. We also note that the binning of the  $\eta\bar{K}^0 $ mass distributions in Fig.~4 in Ref.~\cite{Belle:2022pwd} should be  the same as the one of the $p\bar{K}^0 $   mass distribution to have the same integrated strength as the other distributions.}, where $\beta$ represents the expected color reduction factor associated with the relative strength of internal versus external emission. The resulting mass distributions for all relevant channels are shown in Fig.\ref{res:results}. In particular, the $\eta p$ invariant mass distribution exhibits two distinct peaks: a narrower one around $1535~\text{MeV}$ and a broader one near $1650~\text{MeV}$, which can be associated with the $N^*(1535)$ and $N^*(1650)$ resonances, respectively. These features already emerge from the model without fitting, indicating that the basic framework incorporating final-state interactions among coupled meson-baryon channels captures the essential dynamics responsible for generating these resonant structures.

Yet, we see some difference with the data, in particular, a too small strength of the $N^*(1535)$ state in the $\eta p$ mass distribution and also a small strength  in the $p \bar{K}^0$ mass distribution around 1620~MeV, most likely attributable to the $\Sigma(1620)$ resonance. We will comment on a possible peak around 1490~MeV.

\subsection{Fitting Strategies and Parameter Determination}
We first perform a fit to the Belle data for the $\eta p$ invariant mass distribution, ${d\Gamma}/{dM_\text{inv}(\eta p)}$, to improve the agreement with experimental observations. To better account for the role of the $\pi N$ channel, which lies far away from the $N^*(1535)$ and $N^*(1650)$ resonance regions, we reduce the strength of the $g_{\pi N}$ coupling by 30\%. This adjustment reflects the suppressed contribution of the $\pi N$ channel to the final-state interaction dynamics in the relevant energy range and leads to a more realistic description of the hadronization process.

In addition to this corrections demanded by the data we also introduce a contribution of the $\Sigma(1620)$ resonance. The $\Sigma(1620)$ appears dynamically generated from the $\bar KN,~\pi\Sigma,~\pi\Lambda,~\eta\Sigma,~K\Xi$ coupled channels in~\cite{Meissner:1999vr,Jido:2003cb}. This state is, however, not reproduced very accurately since it appears around 1580~MeV and the width is very large, of the order of 530~MeV in~\cite{Jido:2003cb}, and around 1490~MeV in~\cite{Meissner:1999vr} with a width of about 230~MeV. The couplings obtained for this resonance to the $\bar KN$ channel are also different, $g_{\Sigma^*,~ \bar KN}^{(I=1)}=-1.1-i~1.1$ in~\cite{Jido:2003cb} and $g_{\Sigma^*,~ \bar KN}^{(I=1)}=-0.89-i~0.57$   in~\cite{Meissner:1999vr}. In view of these uncertainties, we let the experiment decide which coupling is required to obtain the strength of the peak of the $\bar K^0p$ mass distribution. For this purpose we parametrize the $\bar K^0p \to \bar K^0p$ amplitude as
\begin{align}\label{newtermKp}
    t_{\bar{K}^0 p,\, \bar{K}^0 p} \equiv \frac{(g_{\Sigma^*,\, \bar{K} N}^{(I=1)})^2}{M^2_\text{inv}(\bar{K}^0 p) - M_{\Sigma^*} + i\, \frac{\Gamma_{\Sigma^*}}{2}}
\end{align}
and we take $M_{\Sigma^*},~\Gamma_{\Sigma^*}$ from the PDG data tables, fitting $g_{\Sigma^*,~ \bar KN}^{(I=1)}$    to reproduce the data.

\begin{figure}[H]
  \centering
 \includegraphics[width=0.33\textwidth]{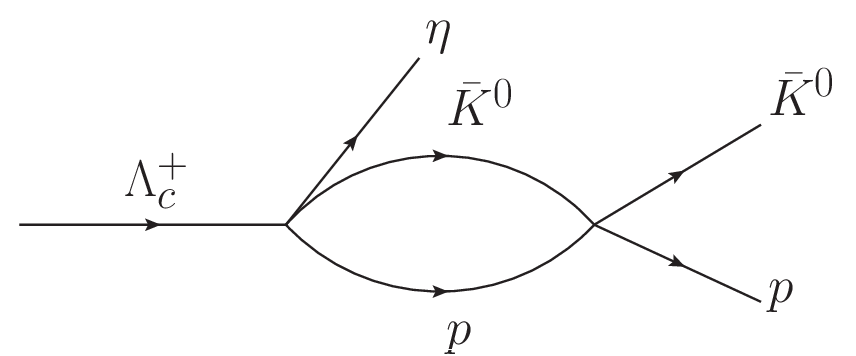}
  \caption{Mechanism to account for the $\Sigma(1620)$  contribution.}
  \label{newterfeymanndigram}
\end{figure}

The contribution of this resonance to the process occurs via final state interaction starting from the $\bar{K}^0 \eta p$  state generated by internal emission in Eq.~(\ref{eq:pb_final}). We have there the term $\frac{1}{\sqrt{6}}~\bar{K}^0 \eta p$, and then the $\bar{K}^0  p$ interacts as shown in Fig.~\ref{newterfeymanndigram}.
Thus, analytically, we add a term   
\begin{align}\label{newtermamp}
        \frac{1}{\sqrt{6}}~G_{\bar{K}^0 p}(M_\text{inv}(\bar{K}^0 p))t_{\bar{K}^0 p,~\bar{K}^0 p}(M_\text{inv}(\bar{K}^0 p))
\end{align}
to Eq.~(\ref{inamp}).
As in~\cite{Meissner:1999vr,Jido:2003cb}, we take $g_{\Sigma^*,~ \bar KN}^{(I=1)}$ complex and find a fair agreement with the data with a value $g_{\Sigma^*,~ \bar KN}^{(I=1)}\sim i~0.6$, qualitatively in agreement with the values obtained in~\cite{Meissner:1999vr,Jido:2003cb}.

The corresponding fitting parameters, $\mathcal{C}$ and $\beta$, and $g_{\Sigma^*,~ \bar KN}^{(I=1)}$ along with the fit quality $\chi^2/\text{d.o.f.}$, are summarized in Table~\ref{tab:results_Parameters}.

\begin{table}[H]
\centering
\caption{Parameters and $\chi^2/\text{d.o.f.}$ for the fit, fixing $g_{\pi N}$ of Table~\ref{tab:weights} with a reduction of 30\%.}
\label{tab:results_Parameters}
\setlength{\tabcolsep}{48pt}
\begin{tabular}{ccc}
\hline \hline
 $\mathcal{C}$ & $\beta$ & $\chi^2/\text{d.o.f.}$ \\
\hline
  $8.01  $ & $0.31 $ & $2.0 $ \\  
\hline \hline
\end{tabular}
\end{table}

The parameter $\beta$, associated with the relative strength of external emission, remains close to the expected color suppression factor $\beta \sim 1/N_c = 1/3$, indicating consistency with the color dynamics of weak decays. 

\subsection{Invariant Mass Distributions}

\begin{figure}[H]
  \centering
 \includegraphics[width=0.33\textwidth]{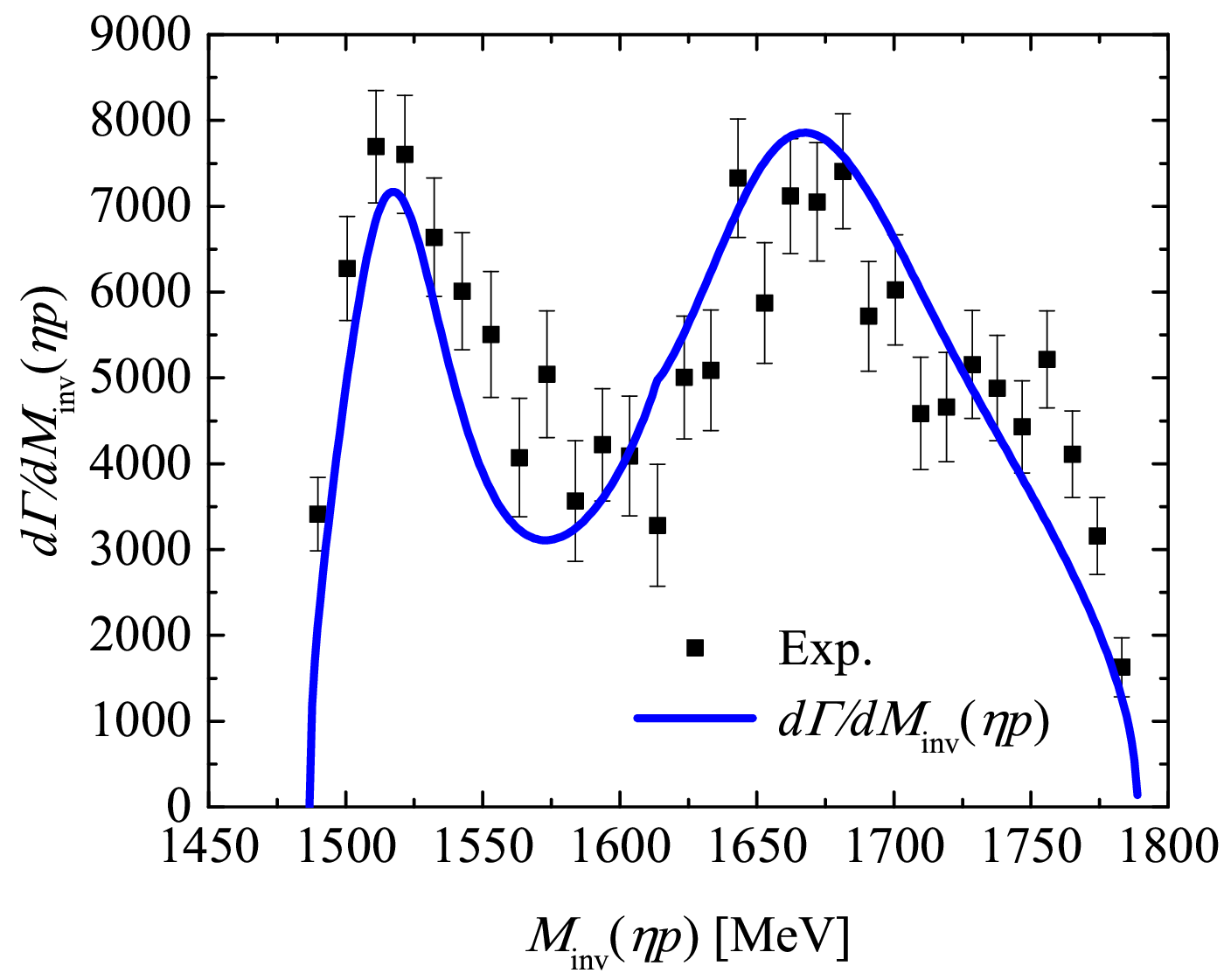}
 \includegraphics[width=0.33\textwidth]{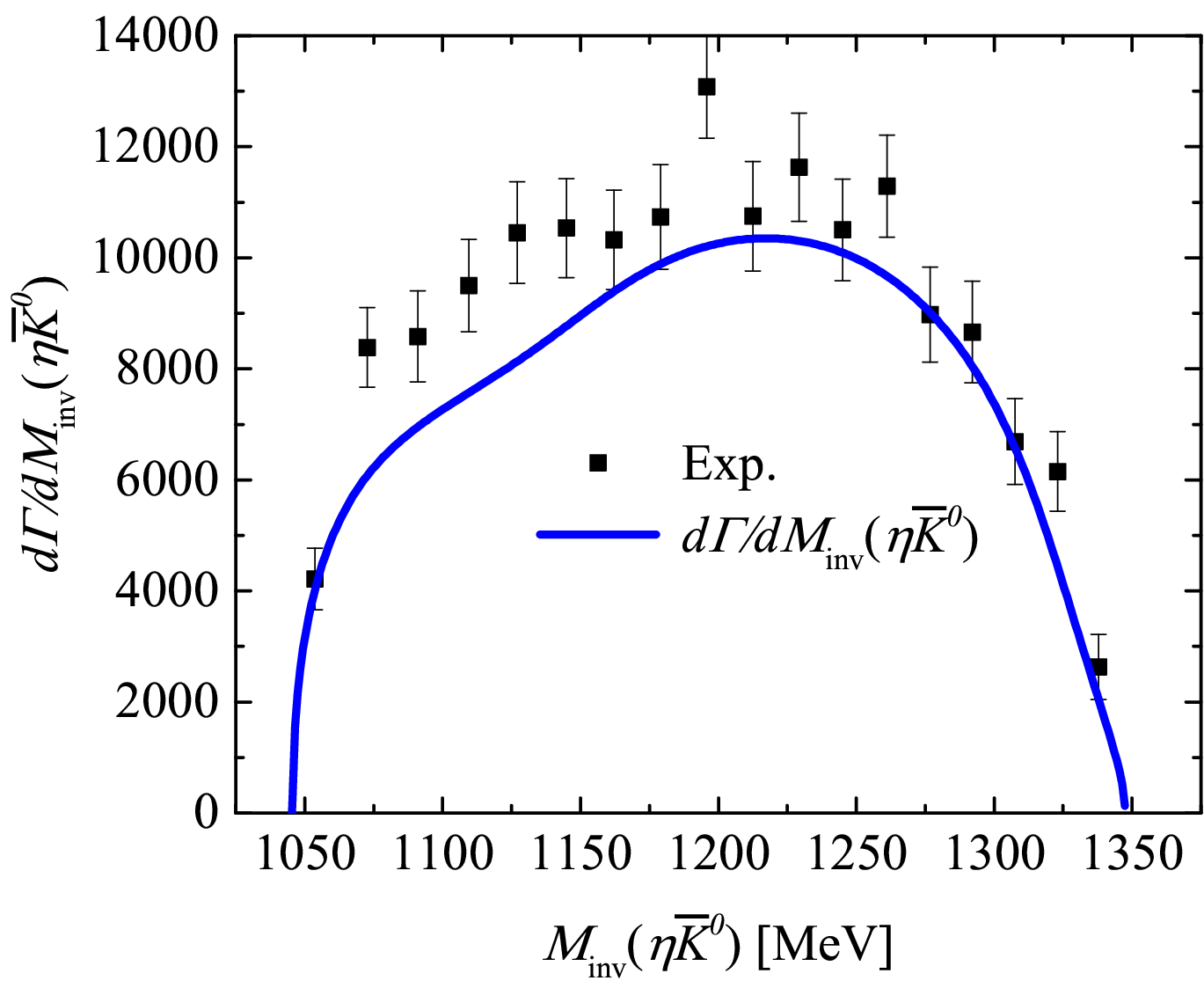}
 \includegraphics[width=0.33\textwidth]{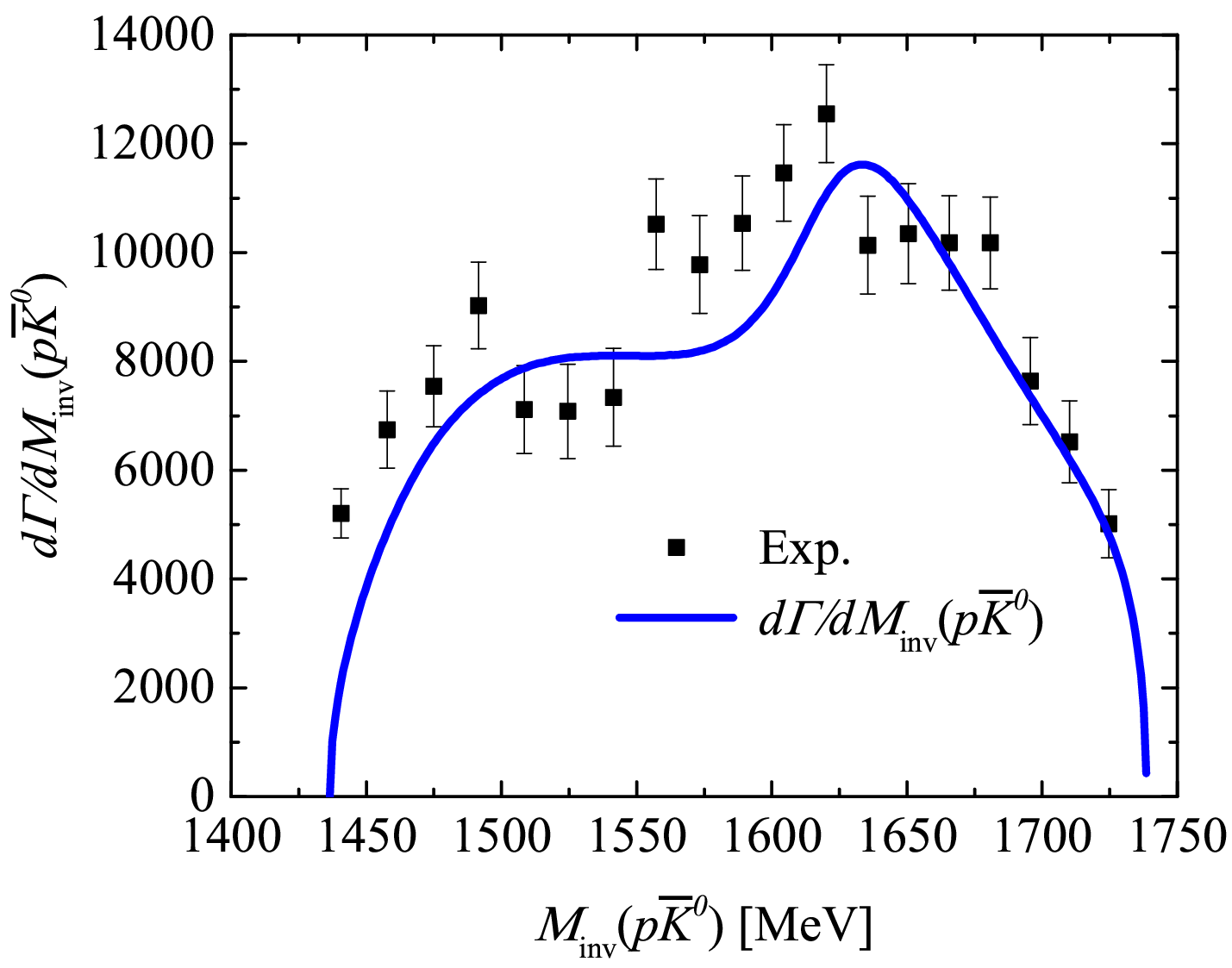}
  \caption{Invariant mass distributions of $\eta p$ (left), $\eta \bar{K}^0$ (middle), and $p \bar{K}^0$ (right). Experimental data is from Belle~\cite{Belle:2022pwd}.}
  \label{res:results_fig1}
\end{figure}

Based on the parameters in Table~\ref{tab:results_Parameters}, we compute the invariant mass distributions ${d\Gamma}/{dM_\text{inv}(\eta p)}$, ${d\Gamma}/{dM_\text{inv}(\eta \bar K^0)}$, and ${d\Gamma}/{dM_\text{inv}(p\bar K^0)}$, and present the results in Fig.~\ref{res:results_fig1}. The mass distributions  are in much better agreement with experimental data than those in Fig.~\ref{res:results}.

We begin with the  discussion of  the $\eta p$ mass distribution ${d\Gamma}/{dM_\text{inv}(\eta p)}$ obtained from these parameters, shown in Fig.~\ref{res:results_fig1} (left). The curve includes both tree-level contributions and the final-state interaction among the coupled meson-baryon channels. Two peaks are clearly observed: a narrower one around $1535~\text{MeV}$ and a broader one around $1650~\text{MeV}$, corresponding to the $N^*(1535)$ and $N^*(1650)$ resonances, respectively.

We then analyze the $\bar{K}^0 p$ mass distribution ${d\Gamma}/{dM_\text{inv}(\bar{K}^0 p)}$ as shown in Fig.\ref{res:results_fig1} (right). The curve includes both the tree-level contributions and the final-state interactions among the relevant meson-baryon coupled channels. A clear peak is observed around $1620~\text{MeV}$, which corresponds to the $\Sigma(1620)$ resonance with $J^P = 1/2^-$.
The $\eta \bar{K}^0$ mass distribution (middle) is rather structureless as also seen in the experiment. 
Coming back to the  $p \bar{K}^0$  mass distribution, the fit produced leaves part of the data points around 1480~MeV above the curve and part of them below, but follows the trend of the bulk of the data. Looking at in from this perspective, and given the absence of any $\Sigma$ resonance in this region, we tend to suggest that the discrepancies in the figure are due to statistical fluctuations of the data, but this is something to be decided by future data with more precision.

\subsection{Impact of Tree-Level Contributions}

\begin{figure}[H]
  \centering
\includegraphics[width=0.45\textwidth]{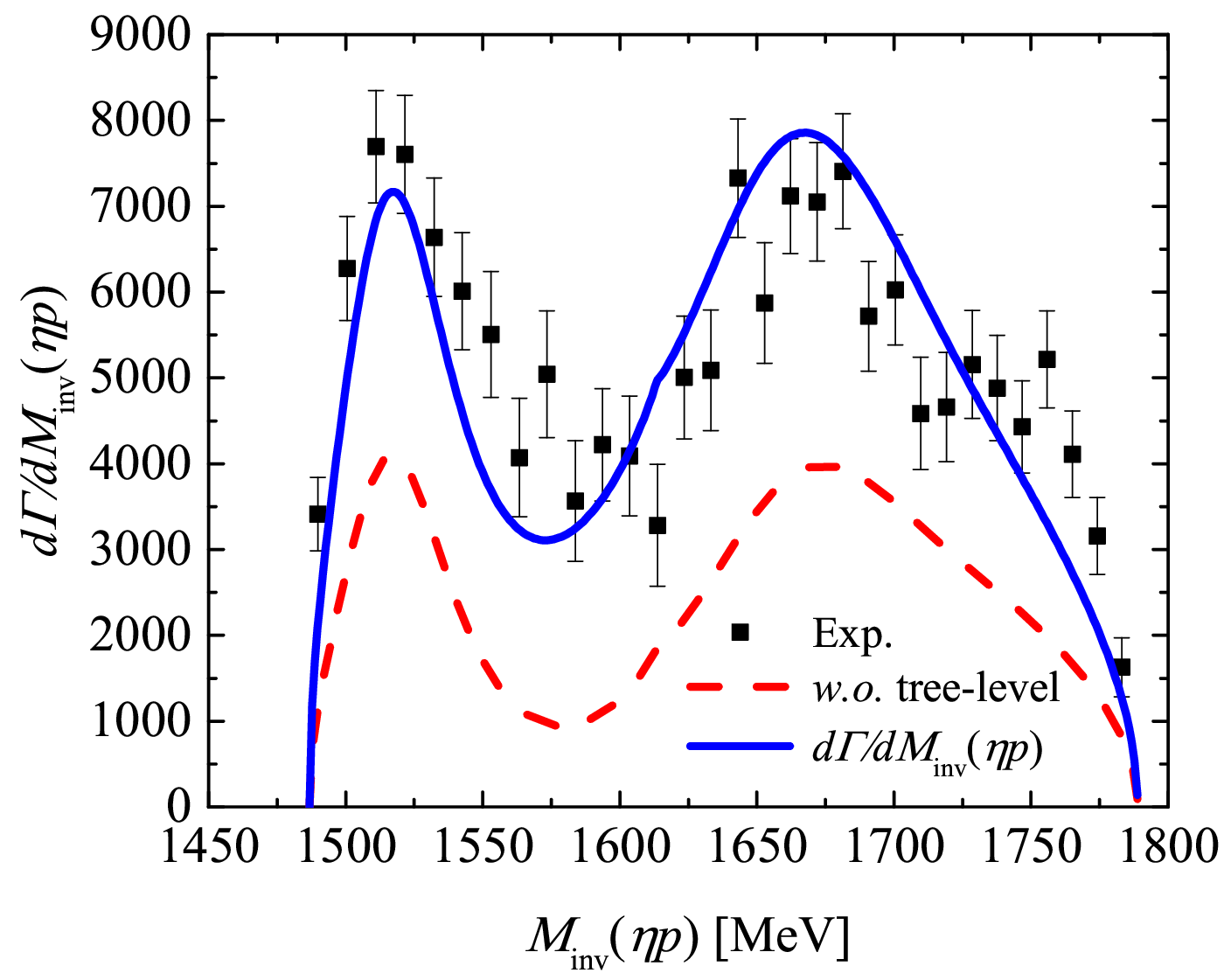}
  \caption{Invariant mass distributions of $\eta p$ with (blue solid) and without (red dashed) the tree-level internal emission contribution. The label 'w.o.' stands for results without tree level.}
  \label{res:results_fig1_wotree}
\end{figure}

To further understand the mechanism behind the resonance generation, we remove the tree-level contribution originating from internal emission and reevaluate the $\eta p$ invariant mass distribution. The results are shown in Fig.~\ref{res:results_fig1_wotree}. Both $N^*(1535)$ and $N^*(1650)$ structures remain clearly visible even after the tree-level contribution is removed, indicating their dynamical origin. This confirms our theoretical assertion that the external emission mechanism, followed by meson-baryon final state interaction, plays a dominant role in the formation of the observed $N^*(1535)$ and $N^*(1650)$ structures. While internal emission directly produces $\bar{K}^0 \eta p$ at tree level, it does not lead to strong $\eta p$ correlations. In contrast, external emission, although not producing $\eta p$ directly, induces rescattering that generates resonant states through coupled-channel dynamics. This highlights the predictive power of our approach, in which some resonances naturally emerge from unitarized meson-baryon dynamics.
We have also checked that without the external emission contribution, the $N^*(1650)$ is not produced in this reaction, in agreement with the results of Refs.~\cite{Xie:2017erh,Pavao:2018wdf}. The external emission mechanism, absent in Ref.~\cite{Xie:2017erh,Pavao:2018wdf} is, thus, essential to produce the $N^*(1650)$ state. 
It is also interesting to see that the contribution of the tree-level is sizeable and, unlike in other reactions where the coherent sum of tree level and resonance contributions produces large interferences~\cite{Song:2025ofe}, in the present case the results of the figure indicate a mostly  incoherent sum.

\subsection{$\Sigma(1620)$ Contributions}

\begin{figure}[H]
  \centering
\includegraphics[width=0.45\textwidth]{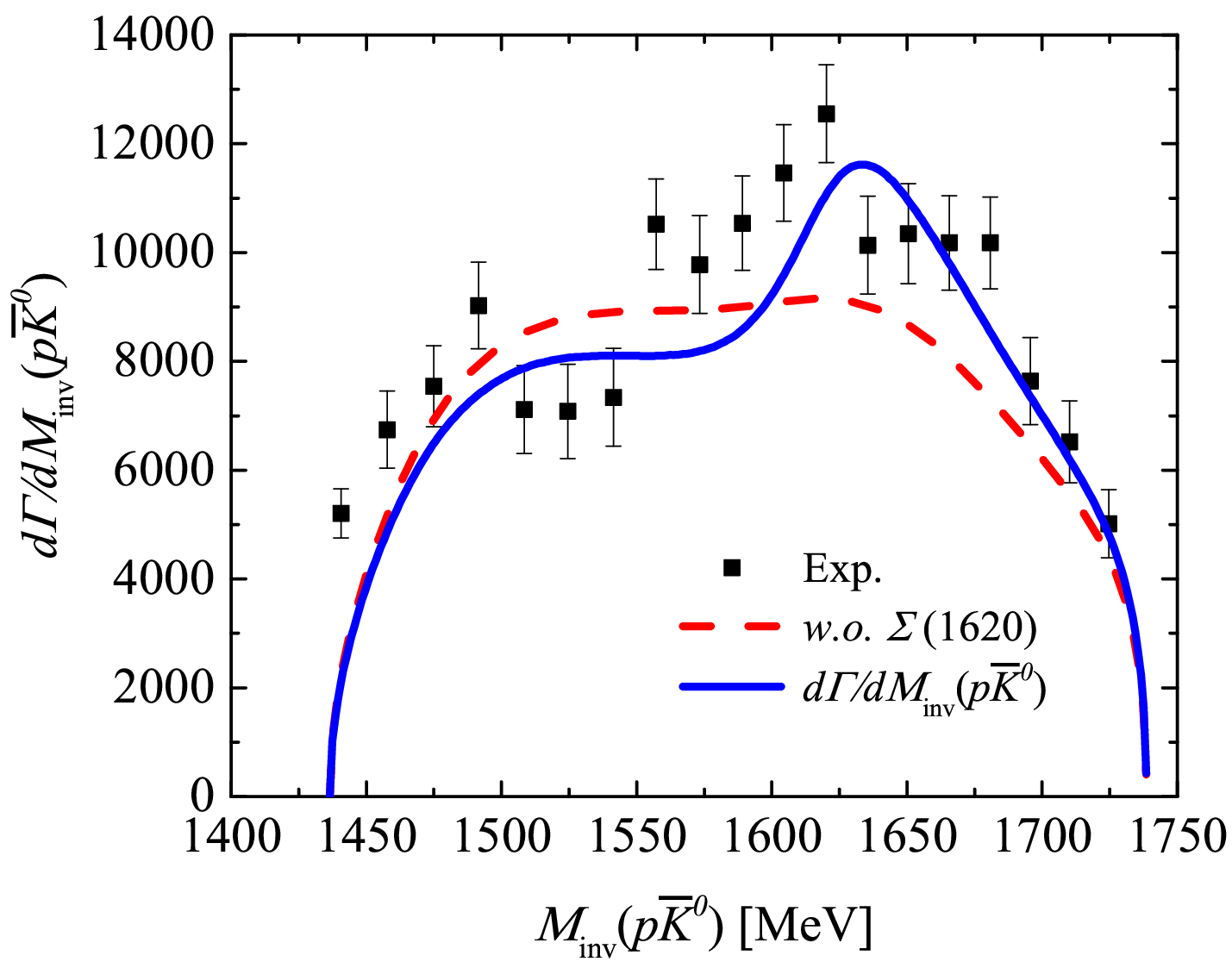}
  \caption{Invariant mass distributions of $\bar{K}^0 p$ with (blue solid) and without (red dashed) the $\Sigma(1620)$ ($J^P = 1/2^-$) contribution. The label 'w.o.' stands for the case without the $\Sigma(1620)$ resonance.}
  \label{res:results_fig2_woSigma}
\end{figure}

To assess the role of the $\Sigma(1620)$ resonance in the $\bar{K}^0 p$ mass distribution ${d\Gamma}/{dM_\text{inv}(\bar{K}^0 p)}$, we remove its contribution from the meson-baryon scattering amplitude and reevaluate the invariant mass distribution. The results are shown in Fig.\ref{res:results_fig2_woSigma}. Compared to the full contribution, the peak around $1620~\text{MeV}$ disappears entirely once the $\Sigma(1620)$ resonance with $J^P = 1/2^-$ is excluded, thus, showing that the peak reflects the 
 $\Sigma(1620)$ contribution. 
This demonstrates that the observed structure is directly linked to the presence of the $\Sigma(1620)$ state. Unlike contact terms or non-resonant background contributions, which depend smoothly on the energy, the disappearance of the peak highlights the resonant nature of the enhancement. 

The results obtained deserve some comment. We have not introduced the $\Sigma(1620)$ empirically as a resonant term to be added to the total amplitude. In our scheme the $\Sigma(1620)$ has appeared through the $\bar K^0p \to \bar K^0p$  rescattering, implicitly assuming that it is a resonance dynamically generated from the interaction of coupled channels. In the contribution if this term we have the weight $\frac{1}{\sqrt{6}}$ of $\bar K^0 \eta p$ production and the factor $\beta$ of internal emission of Eq.~(\ref{inamp}), the $G_{\bar K^0p}$  function and the scattering matrix of  Eq.~(\ref{newtermamp}). With all these factors the empirical coupling $g_{\Sigma^*,\, \bar{K} N}^{(I=1)}$  obtained from a fit to the data is in qualitative agreement with the results of~\cite{Meissner:1999vr,Jido:2003cb}.
This provides additional support for the interpretation of $\Sigma(1620)$ as a dynamically generated state within the unitarized coupled-channel framework in spite of present theoretical uncertainties. These findings should also serve as an incentive to improve present models that go  beyond the lowest orders chiral Lagrangians~\cite{Ramos:2016odk,Feijoo:2021zau,Krause:1990xc,Frink:2004ic}, and which produce this resonance, including these additional constraints. 

\subsection{Discussions}
We present the $\eta p$, $\bar{K}^0 p$, and $\bar{K}^0 \eta$ mass distributions for the decay $\Lambda_c^+ \to \bar{K}^0 \eta p $, showing that the $N^*(1535)$, $N^*(1650)$, and $\Sigma(1620)$ resonances naturally emerge from final-state interactions in the chiral unitary approach. Even without parameter tuning, the model successfully reproduces the main resonant structures, demonstrating its predictive power.
Moreover, our fitting procedure reinforces this picture: moderate reductions in the $g_{\pi N}$ coupling improve the agreement with the Belle data and successfully reproduce all key resonance structures in both the $\eta p$ and $\bar{K}^0 p$ channels.

The stability of the $N^*(1535)$ and $N^*(1650)$ resonance peaks, even after removing tree-level contributions, offers compelling evidence that both states emerge dynamically from meson-baryon interactions. Within this approach, the $N^*(1535)$ exhibits strong couplings to $\eta N$ and $K \Lambda$, while the $N^*(1650)$ couples predominantly to $\pi N$ and $\rho N$. These distinct coupling patterns manifest in the $\eta p$ invariant mass distribution, further supporting their interpretation as dynamically generated or hadronic molecular states.
Moreover, the robustness of the resonance signals after removing the tree-level contribution confirms the dominant role of external emission mechanisms followed by final-state interactions in weak $\Lambda_c^+$ decays. In particular, the appearance of the peak around $1620~\text{MeV}$ in the $\bar{K}^0 p$ invariant mass distribution in the present framework, as a consequence of final state interaction and with couplings of the resonance to $\bar K N~(I=1)$ similar to those obtained in the chiral unitary approach, further reinforces the pictures of the  $\Sigma(1620)$ as a dynamically generated state. These findings highlight the predictive power of the unitarized coupled-channel meson-baryon framework in describing the emergence of hadronic molecular states from final-state interactions.

These analyses demonstrate that the $\Lambda_c^+ \to \bar{K}^0 \eta p $ reaction provides a valuable tool for investigating the nature of $N^*$ resonances, particularly in exploring their possible interpretation as dynamically generated states from meson-baryon coupled-channel interactions. Remarkably, this study presents the first clear theoretical demonstration of the simultaneous appearance of two related $N^*$ resonances—$N^*(1535)$ and $N^*(1650)$—in this decay process. Additionally, the appreciable contribution from the $\Sigma(1620)$ resonance in the $\bar{K}^0 p$ invariant mass distribution provides further support for its interpretation as a hadronic molecular state emerging from the coupled-channel dynamics inherent to the chiral unitary approach.

The decay $\Lambda_c^+ \to \bar{K}^0 \eta p$ offers a particularly clean environment to probe such structures. The process is dominated by an external emission topology followed by meson-baryon rescattering, making it highly sensitive to the $I=1/2$ meson-baryon interaction dynamics. 
Future experimental studies with improved resolution of the $\eta p$ and $\bar{K}^0 p$ invariant mass spectrum in $\Lambda_c^+$ decays could offer precise constraints on the properties and internal structure of the $N^*(1535)$, $N^*(1650)$, and $\Sigma(1620)$. Such experimental measurements would be instrumental in testing coupled-channel predictions and in disentangling the molecular versus compact quark core components of these resonances.

\section{Conclusion} \label{sec:Conclusion}

In this work, we have revisited the decay process $\Lambda_c^+ \rightarrow \bar{K}^0 \eta p$ with a comprehensive theoretical framework that incorporates both internal and external weak emission mechanisms, as well as final-state interactions within a coupled-channel chiral unitary approach. By including both pseudoscalar-baryon and vector-baryon channels, our model successfully generates the $N^*(1535)$, $N^*(1650)$, and $\Sigma(1620)$ resonances dynamically, which appear as distinct and relatively close structures in the $\eta p$ and $\bar{K}^0 p$ invariant mass distributions.

Our study reveals that the $N^*(1535)$ and $N^*(1650)$ resonances observed in the $\eta p$ final state are naturally produced through unitarized meson-baryon coupled-channel dynamics, driven predominantly by the external weak emission mechanism. These resonances emerge as poles in the scattering amplitude due to multiple rescatterings among intermediate states such as $\pi N$, $\eta N$, $K\Lambda$, and $\rho N$, all of which couple to the $\eta p$ system. Importantly, we demonstrate that both $N^*$ structures remain clearly visible even after removing the tree-level of internal emission contribution, confirming their dynamical origin and the dominant role of the external emission mechanism in weak $\Lambda_c^+$ decays.

A similar analysis is carried out for the $\bar{K}^0 p$ invariant mass distribution, where a clear peak is found around $1620~\text{MeV}$, associated with the $\Sigma(1620)$ resonance with $J^P = 1/2^-$. 
We introduce the $\Sigma(1620)$ through final state  interaction of $\bar K^0p \to \bar K^0p$, and we find that the data calls for a coupling of the resonance to the $\bar K N$ channel in line with findings of the chiral unitary approach, providing support for the dynamical origin of this resonance from the meson baryon interaction.

The decay $\Lambda_c^+ \to \bar{K}^0 \eta p$ thus offers a particularly clean environment to probe the nature of such baryon resonances. The dominance of the external emission mechanism, together with the sensitivity of the final-state interaction to $I = 1/2$ meson-baryon dynamics, allows for the selective enhancement of specific resonant contributions. The $\Sigma(1620)$ signal in the $\bar{K}^0 p$ distribution further validates the predictive capability of our model.

Importantly, this is the first time that two related $N^*$ resonances, $N^*(1535)$ and $N^*(1650)$, have been simultaneously observed and clearly distinguished in the meson-baryon final state of this decay channel. Our results demonstrate that both resonances emerge naturally from meson-baryon coupled-channel dynamics, with their production strongly influenced by final state interactions, especially through the external emission mechanism.
This study not only reproduces the main features observed in the experimental $\eta p$ mass distribution  by Belle but also demonstrates a consistent and simultaneous theoretical description of the $N^*(1535)$, $N^*(1650)$, and $\Sigma(1620)$ resonances in a single decay process. These findings support the interpretation of these states as dynamically generated hadronic molecules, thereby shedding new light on the internal structure of low-lying baryon resonances.

In summary, the combined treatment of weak decay mechanisms and unitarized meson-baryon interactions provides a robust theoretical framework to explore resonance formation in charmed baryon decays. The $\Lambda_c^+ \to \bar{K}^0 \eta p$ decay emerges as an excellent probe for such studies, offering new insights into the baryon spectrum in the non-perturbative regime of QCD and opening the door to further investigations in related processes.

\section*{ACKNOWLEDGMENTS}
This work is partly supported by the National Natural Science
Foundation of China under Grants  No. 12405089 and No. 12247108 and
the China Postdoctoral Science Foundation under Grant
No. 2022M720360 and No. 2022M720359.  Yi-Yao Li is supported in part by the Guangdong Provincial international exchange program for outstanding young talents of scientific research in 2024. 
 This work is partly supported by the National Natural Science Foundation of China (NSFC) under Grants No. 12365019 and No. 11975083, and by the Central Government Guidance Funds for Local Scientific and Technological Development, China (No. Guike ZY22096024).
 This work is also partly supported by the Spanish Ministerio de Economia y Competitividad (MINECO) and European FEDER
  funds under Contracts No. PID2020-112777GB-I00.
Raquel Molina acknowledges support from
the ESGENT program (ESGENT/018/2024) and the PROMETEU program (CIPROM/2023/59), of the Generalitat Valenciana, and also
from the Spanish Ministerio de Economia y Competitividad and
European Union (NextGenerationEU/PRTR) by the grant
(CNS2022-13614. This research partially supported by grant PID2023-147458NB-C21 funded by MCIN/AEI/ 10.13039/501100011033 and by the European Union.

\bibliography{refs.bib} 
\end{document}